\documentclass[a4paper,fleqn]{cas-sc}

\usepackage[authoryear,longnamesfirst]{natbib}

\usepackage{multicol}
\usepackage{multirow}
\usepackage{tabularx}
\usepackage{booktabs}
\usepackage{algpseudocode}
\usepackage{subcaption}
\newtheorem{prop}{Proposition}
\usepackage{color}
\usepackage{caption}
\usepackage{float}
\usepackage{adjustbox}

\usepackage{tikz}
\usetikzlibrary{arrows,automata,backgrounds,fit,patterns,petri,shadows,shapes,calc}
\pgfdeclarepatternformonly{stripes}
{\pgfpointorigin}{\pgfpoint{0.25cm}{0.25cm}}
{\pgfpoint{0.25cm}{0.25cm}}
{
    \pgfpathmoveto{\pgfpoint{0cm}{0cm}}
    \pgfpathlineto{\pgfpoint{0.25cm}{0.25cm}}
    \pgfpathlineto{\pgfpoint{0.25cm}{0.12cm}}
    \pgfpathlineto{\pgfpoint{0.12cm}{0cm}}
    \pgfpathclose%
    \pgfusepath{fill}
    \pgfpathmoveto{\pgfpoint{0cm}{0.12cm}}
    \pgfpathlineto{\pgfpoint{0cm}{0.25cm}}
    \pgfpathlineto{\pgfpoint{0.12cm}{0.25cm}}
    \pgfpathclose
    \pgfusepath{fill}
}
\tikzstyle{decisionA} = [regular polygon, regular polygon sides = 4, thick, minimum size = 1.0cm, draw = black, fill = gray!40]
\tikzstyle{decisionD} = [regular polygon, regular polygon sides = 4, thick, minimum size = 1.0cm, draw = black]
\tikzstyle{utilityA} = [regular polygon, regular polygon sides = 6, thick, minimum size = 4cm, draw = black, fill = gray!20]
\tikzstyle{utilityD} = [regular polygon, regular polygon sides = 6, thick, minimum size = 4.0cm, draw = black]
\tikzstyle{chanceA} = [circle, thick, minimum size = 1.0cm, inner sep = 0.1pt, draw = black, fill = gray!40]
\tikzstyle{chanceD} = [circle, thick, minimum size = 1.0cm, inner sep = 0.1pt, draw = black]
\tikzstyle{chance} = [circle, thick, minimum size = 1.0cm, inner sep = 0.1pt, draw = black, pattern = stripes, pattern color = gray!40]
\tikzstyle{empty} = [circle, line width = 0pt, minimum size = 1cm, inner sep = 0.1pt]

\usepackage{graphicx}
\usepackage[space]{grffile}
\usepackage{latexsym}
\usepackage{textcomp}
\usepackage{longtable}
\usepackage{tabulary}
\usepackage{booktabs,array,multirow}
\usepackage{amsfonts,amsmath,amssymb}
\usepackage{enumitem}
\providecommand\citet{\cite}
\providecommand\citep{\cite}

\newif\iflatexml\latexmlfalse

\AtBeginDocument{\DeclareGraphicsExtensions{.pdf,.PDF,.eps,.EPS,.png,.PNG,.tif,.TIF,.jpg,.JPG,.jpeg,.JPEG}}

\usepackage[utf8]{inputenc}
\usepackage[english]{babel}

\newcommand{\bc}{\begin{center}}
\newcommand{\ec}{\end{center}}
\newcommand{\Pb}{\hbox{{I}\kern-.1667em\hbox{P}}}
\newcommand{\Ex}{\hbox{{I}\kern-.1667em\hbox{E}}}
\newcommand{\Real}{\hbox{{I}\kern-.1667em\hbox{R}}}
\newcommand{\sReal}{\hbox{\scriptsize {I}\kern-.1667em\hbox{R}}}
\def\tsc#1{\csdef{#1}{\textsc{\lowercase{#1}}\xspace}}
\tsc{WGM}
\tsc{QE}

\newcommand{\Dcal}{\mathcal{D}}

\newcommand{\dd}{\mathop{}\! \mathrm{d}}
\newcommand{\argmax}{\operatorname*{\arg\max}}
\newcommand{\argmin}{\operatorname*{\arg\min}}

\expandafter\def\csname __first_footerline:\endcsname{}
\begin{document}
\let\WriteBookmarks\relax

\shorttitle{}    
\shortauthors{}  

\title [mode = title]{Adversarial risk analysis with fully probabilistic designs}  



%

\author[1,2]{Mario Chacón-Falcón}

\cormark[1]


\ead{mario.chacon@icmat.es}

\ead[url]{https://mchacon28.github.io/}

\credit{Conceptualization, Methodology, Formal analysis, Software, Writing - Original Draft, Writing - Review \& Editing}

\affiliation[1]{organization={Institute of Mathematical Sciences (ICMAT-CSIC)},
            city={Madrid},
            postcode={28049}, 
            country={Spain}}
\affiliation[2]{organization={Universidad Autónoma de Madrid, Escuela de Doctorado},
            city={Madrid},
            postcode={28049}, 
            country={Spain}}

\author[3]{Tatiana Valentine Guy}


\ead{guy@utia.cas.cz}

\ead[url]{https://utia.cas.cz/en/people/?pid=75}

\credit{Conceptualization, Methodology, Formal analysis, Writing - Review \& Editing}
\author[3]{Miroslav Kárný}


\ead{school@utia.cas.cz}

\ead[url]{https://utia.cas.cz/en/people/?pid=109}
\affiliation[3]{organization={Dept. Adaptive Systems, UTIA, Czech Academy of Sciences},
            city={Prague},
            country={Czech Republic}}

\credit{Conceptualization, Methodology, Formal analysis, Writing - Review \& Editing}

\author[1]{David Rios Insua}



\ead{david.rios@icmat.es}

\ead[url]{https://www.davidriosinsua.es/}

\credit{Conceptualization, Methodology, Formal analysis, Writing - Review \& Editing, Supervision}




\begin{abstract}
 Adversarial risk analysis (ARA) informs decision-makers when facing intelligent opponents and uncertain outcomes, enabling an analyst to model beliefs about the opponents' preferences, capabilities, beliefs, and the strategic calculations that opponents employ. Major applications cover reliability and risk analysis of critical infrastructures in presence of adversarial threats.
In such domains, ARA mitigates typical common knowledge assumptions of standard game-theoretic approaches. However, ARA entails complex modelling and computations. To alleviate these, we explore its combination with fully probabilistic design (FPD), a control theory methodology supporting an agent in choosing the decision distribution that achieves the outcomes with a distribution closest to its ideal. We use FPD to model and forecast the opponent's decisions within an ARA setting and present how to handle security games within this ARA-FPD framework, covering both single and multiple stage problems.
\end{abstract}



\begin{keywords}
Decision analysis \sep Adversarial Risk Analysis \sep Security \sep Fully Probabilistic Design \sep Reliability Analysis
\end{keywords}

\maketitle

\section{Introduction}
\noindent
Numerous reliability and risk analysis settings, specially when dealing with critical infrastructures, comprise decisions
by adversaries affecting the security, safety and acceptability of systems, see the reviews by \citet{hausken2025review} and \citet{guikema2010assessing}, as well as references therein.
Examples in reliability include 
 life testing problems  \citep{ReliabilityLifeSciences}; warranty analysis \citep{ReliabilityWarranty}; optimal maintenance \citep{ReliabilityManteinance}; reliability demonstration \citep{RUFOReliability}; and software reliability 
 testing  \citep{soyer2025adversarial}. Examples in risk analysis include transportation network protection \citep{quijano2018critical}, nuclear power power plant protection \citep{zhao2020finite} or counterterrorism 
 protection of complex infrastructures \citep{hausken2010defense}.
 
 Decision making when facing intelligent opponents with non-aligned or misaligned objectives is frequently formulated within standard game-theoretic perspectives which, from a computational point of view, are based on approximating Nash equilibria and related refinements \citep{menache2011network}.
The underlying assumptions to these perspectives usually entail that agents know not only their payoffs, preferences, beliefs, and feasible actions, but also those of their opponents, usually designated {\em common knowledge} (CK) assumptions.
Alternatively, in games with incomplete information \citep{Harsanyi1967} it is habitually assumed that each agent has a probability distribution over certain opponents' features, designated types, characterizing their beliefs and preferences.
 This distribution is considered to be known by all players, a specific 
 CK assumption called the {\em common prior } hypothesis, allowing for a symmetric joint normative analysis of the game in which players maximize expected utilities, and expect other players to do the same, leading to Bayes-Nash equilibria.  General CK assumptions have been critically reviewed, e.g., in \citet{Raiffa2002} or \citet{Hargreaves:2004}, whereas common prior assumptions were discussed by \cite{antos2010representing}, \cite{sakovics2001games} or \cite{angeletos2018forward}, to name but a few. In their reviews of game theoretic approaches to reliability and risk analysis,
 \citet{hausken2025review} and \citet{hunt2024review} argue that developing methodology addressing cases with lack of full knowledge and full rationality are
 important topics for further development.

Adversarial risk analysis (ARA)  \citep{RiosInsua2009,banks2015adversarial} is a decision-analytic approach to games that employs distributions to model the goals, resources, beliefs, preferences, and reasoning of an attacker precisely aiming to address such lack of full knowledge and rationality issues.
Within ARA, a defender solves the problem from the perspective of her attacker, while placing subjective probability distributions over all the attacker's unknown problem elements.  This gives her a distribution over the attacker's actions, accounting for the uncertainty she may have about her adversary's actions, enabling her to maximize expected utility to find her optimal protective action. 
\citet{Banketal2022} provide a detailed conceptual comparison between ARA and key game theoretic solutions, including Nash and Bayes-Nash equilibria. 
Since its proposal, ARA has been used to model a variety of problems, mainly in the security risk domain, including routing in the presence of improvised explosive devices \citep{wang11}, international piracy \citep{sevillano12},  counter-terrorist online surveillance \citep{Gil19}, enhanced combat modelling \citep{roponen2020adversarial}, and cybersecurity \citep{wang2021bayesiannetworkbasedcybersecurityadversarialrisk}. Reliability applications include \cite{RIOSINSUAReliability} and \citet{GonzalezOrtega2021}. 
However, ARA entails involved modelling and computational issues. In particular, to account for the uncertainty about the attacker, sophisticated probability distributions need to be assessed, including random utilities, 
 demanding complex reasoning. Furthermore, the ARA methodology entails simulations from the attacker's perspective and then optimization of the defender's problem, potentially leading to expensive computations.

This paper studies how these issues are alleviated with the help of the control-theoretic methodology of Fully Probabilistic Design (FPD) \citep{KARNYFPD2006}, introducing a novel ARA-FPD approach. Within stochastic control problems, FPD of decision rules supports the decision-making of an agent while expressing its decision objectives through an ideal distribution that describes the consequences sought by such agent \citep{KARNY2022}. 
Here, FPD will be used to facilitate forecasting the attacker's actions within ARA settings, greatly alleviating the aforementioned modelling and computational issues by, on the one hand, mitigating the need to extract very detailed information about the attacker's preferences, limiting the focus to his ideals, and, on the other hand, avoiding repeated costly optimizations adapting FPD explicit optimization results. 

For this, Section \ref{sec:Seq_Games} introduces key notation and illustrates basic concepts integrating ARA and FPD through two simple templates concerning sequential and simultaneous defend-attack interactions as well as comparing ARA-FPD with ARA and game theoretic  approaches. Then, an ARA-FPD approach to multi-stage interactions is introduced in Section \ref{sec:Rep_games}, including an approach to sequentially learn attacker's goals. Section \ref{sec:Exp} provides empirical results on the effectiveness of the proposed methodology.

Table \ref{tab:not_summary} contains key notation. Proof of propositions can be found in Appendix A.
The site \url{https://github.com/mchacon28/ARAFPD} contains code to reproduce the examples. 
\begin{table*}[pos =h]
\centering
\caption{Key notation}
\begin{adjustbox}{width=0.6\columnwidth}
\begin{tabular}{c|c}
\hline
\textbf{Notation} & \textbf{Concept} \\ \hline
$D$ & \begin{tabular}[c]{@{}c@{}}Defender\end{tabular} \\ \hline
$A$ & \begin{tabular}[c]{@{}c@{}}Attacker \end{tabular} \\ \hline
$\mathcal{D}$ & 
\begin{tabular}[c]{@{}c@{}}Set of feasible defences 
\end{tabular} \\ \hline
$\mathcal{A}$ & Set of feasible attacks  \\ \hline
$\Theta$ & \begin{tabular}[c]{@{}c@{}} Set of possible outcomes\end{tabular} \\ \hline
$p_D(\theta \vert a,d)$ & \begin{tabular}[c]{@{}c@{}} Density modelling defender's beliefs over outcomes \end{tabular} \\ \hline
$p_A(\theta \vert a,d)$ & \begin{tabular}[c]{@{}c@{}} Density modelling attacker's beliefs over outcomes  \end{tabular} \\ \hline
$u_D$ & Defender's utility function\\ \hline
$u_A$ & Attacker's utility function\\ \hline
$\psi_D$ & Defender's expected utility \\ \hline
$\psi_A$ & Attacker's expected utility \\ \hline
$U_A$ & Random variable modelling $D$'s beliefs about $u_A$ \\ \hline
$P_A$ & Random variable modelling $D$'s beliefs about $p_A$ \\ \hline
$\Psi_A$ & Random variable modelling $D$'s beliefs about $\psi_A$ \\ \hline
$p_D(a\vert d)$ &  Model of $D$'s beliefs about $A$'s attack given her defence $d$\\ \hline
$p_A(d)$ & Density modelling $A$'s beliefs about $D$'s defence \\ \hline
$\Pi_A$ & Random variable modelling $D$'s beliefs about $p_A(d)$ \\ \hline
$d^*$ & Defender's optimal defence \\ \hline
$\mathrm{KL} [\cdot \| \cdot ] $ & Kullback-Leibler divergence \\ \hline
$p^i_A(\theta \vert a,d)$ & $A$'s ideal outcome realization density  given $a$ and $d$\\ \hline
$A^*$ & Model of $D$'s beliefs about $A$'s optimal attack \\ \hline
$P^i_A(\theta \vert a,d)$ & Random variable modelling $D$'s beliefs about $p^i_A(\theta \vert a,d)$  \\ \hline
$p^i_A(a \vert d)$ & $A$'s ideal attack density given $D$'s defence $d$ \\ \hline
$P^i_A(a \vert d)$ & Random variable modelling $D$'s beliefs about $p^i_A(a \vert d)$ \\ \hline
$\beta e(\alpha,\beta)$ & Beta distribution with parameters $\alpha$, $\beta$ \\ \hline
$Dir(\alpha_1,\alpha_2,\alpha_3)$ & Dirichlet distribution with parameters $\alpha_1$, $\alpha_2$, $\alpha_3$ \\ \hline

\end{tabular}
\end{adjustbox}
\label{tab:not_summary}
\end{table*}
\section{ARA-FPD for single-stage defend-attack interactions} \label{sec:Seq_Games}

\subsection{Sequential defend-attack games} \label{sec:Seq_gamesARAreview}
\begin{figure}[pos= h]
\centering
\begin{tikzpicture}[->,>=stealth',shorten >=1pt,auto,node distance=2.5cm, scale=0.9, transform shape, semithick]

  \tikzstyle{decisionA} = [rectangle, thick, minimum width=1.6cm, minimum height=0.8cm, draw=black, fill=gray!40, align=center]
  \tikzstyle{decisionD} = [rectangle, thick, minimum width=1.6cm, minimum height=0.8cm, draw=black, align=center]

  \tikzstyle{utilityA} = [regular polygon, regular polygon sides=6, thick, minimum size=1.7cm, draw=black, fill=gray!40, align=center, inner sep=1pt, text width=0.8cm, font=\scriptsize]
  \tikzstyle{utilityD} = [regular polygon, regular polygon sides=6, thick, minimum size=1.7cm, draw=black, align=center, inner sep=1pt, text width=0.8cm, font=\scriptsize]

  \tikzstyle{chance} = [circle, thick, minimum size=0.6cm, inner sep=1pt, draw=black, pattern=stripes, pattern color=gray!40, align=center, text width=1.4cm, font=\small]

  \node[chance](D) {Outcome};
  \node[decisionD] (A) [left of=D, xshift=-0.5cm] {Defender};
  \node[decisionA] (B) [right of=D, xshift=0.5cm] {Attacker};
  \node[utilityD]  (C) [below of=A, yshift=-0.2cm] {Defender \\ Utility};
  \node[utilityA]  (E) [below of=B, yshift=-0.2cm] {Attacker \\ Utility};

  \path (A) edge    node {} (D)
            edge    node {} (C)
        (B) edge    node {} (D)
            edge    node {} (E)
        (D) edge    node {} (C)
            edge    node {} (E)
        (A) edge[out=90, in=90, dashed] node {} (B);
\end{tikzpicture}
\caption{Basic two-agent sequential defend-attack game as a BAID}
\label{fig:tpsdg}
\end{figure}
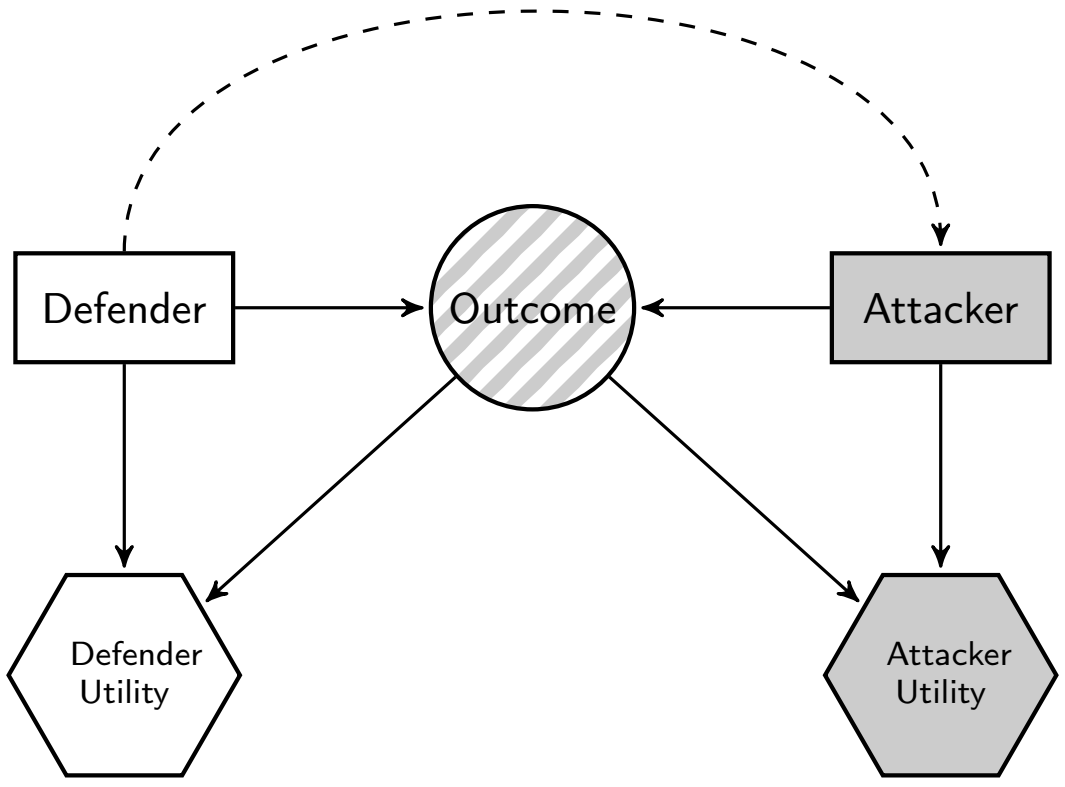

Consider (single-stage) sequential defend-attack games \citep{Brown:2006}, also known as Stackelberg games \citep{Gibbons:1992}. They are represented in Figure \ref{fig:tpsdg} through a bi-agent influence
diagram (BAID).\footnote{Bi-agent influence diagrams (BAIDS) \citep{banks2015adversarial} facilitate visualizing decision problems
  with two agents. Circular nodes represent uncertainties; hexagonal ones, evaluations; and square nodes indicate decisions. Arcs pointing to decision nodes (dashed) represent that such decisions are made knowing the values of the preceding nodes; arcs pointing to chance and utility nodes indicate that the predecessors respectively influence these events or utilities. Coloured nodes indicate relevance to just one of the agents (white, defender; gray, attacker); striped nodes are relevant to both agents.}
A defender ($D$, she) chooses a decision $d \in \Dcal$.  Then,  an attacker ($A$, he), after observing $d$, chooses his attack $a\in {\cal A}$. The consequences for both agents depend on an outcome  $\theta \in \Theta$ resulting from the attack and defence
  interaction.

\noindent Each agent is assumed to have its own beliefs on the possible outcomes, given the applied defence and attack, respectively modelled through the densities $p_D(\theta \vert a,d)$ and $p_A(\theta  \vert a,d)$. Similarly, their utility functions are $u_D(d, \theta)$ and $u_A( a, \theta)$.
   Then, both agents know the expected utility that a pair $(d,a)$ would provide them, 
$ \psi_j (d,a ) = \int u_j (d, \theta ) p_j(\theta | a, d) \dd \theta$, for $j \in \{ D,A\}$.
The game-theoretic solution does not require the attacker to know $D$'s judgments, as he observes her decision, and simply computes his best attack response maximizing his expected utility given the implemented defence.  However, the defender must know those of the attacker, which is the CK assumption in this setting, leading to Nash (and sub-game perfect) equilibria \citep{Hargreaves:2004}.

The ARA approach to sequential games relaxes this CK assumption: the defender is uncertain about the attacker's response to her action $d$ and models this uncertainty through a density $p_{D}(a \vert d)$. Then, adopting a decision analytic approach, her expected utility for implementing her  defence $d$ would be $\psi_{D}(d) = \int_{\mathcal{A}} \psi_D(d,a)\, p_D(a \vert d) \dd a$, her optimal decision being
\begin{equation} \label{kantorei}
    \begin{aligned}
         d^* &=\argmax_{d \in {\cal D}}
           \int_{\mathcal{A}}  \psi_D (d, a) p_D (a \vert d ) \dd a = \argmax_{d\in {\cal D}}
           \int_{\mathcal{A}}  \int_{\Theta}  u_D (d,\theta ) p_D (\theta  \mid a,d) p_D (a \vert d ) \dd\theta \dd a.    
    \end{aligned}
\end{equation}   
 \noindent In (\ref{kantorei}), while $u_D(d, \theta)$ and $p_D(\theta \vert a,d)$ refer to standard decision analytic assessments \citep{clemen2013making}, $p_D(a \vert d)$ entails strategic aspects that need to be taken into account.
    
For this, ARA initially aims to assess $p_D(a \vert d)$ by modelling $A$'s problem as if he maximized his expected utility given the action $d$ he has observed, through \begin{small}
    $a^*(d)= \argmax_{a\in \mathcal{A}} \psi_A (d,a).$
\end{small} However, the defender does not have full access to the utility $u_A$ nor his density $p_A$ \citep{KEENEY}. Suppose she models her uncertainty about these elements through a distribution \begin{small}
    $F = (U_A(a,\theta),P_A(\theta \vert a,d))$
\end{small} of random utilities $U_A(a, \theta)$ and densities $P_A (\theta \vert a,d)$ defined over a common probability space \citep{chung2000course}.  
   This, in turn, 
     induces a distribution over $A$'s optimal action, given by 
     \begin{equation*}
     \begin{aligned}
       A^*(d) &= \argmax_{a\in\mathcal{A}} \Psi_A (d,a) = \argmax_{a\in\mathcal{A}} \int_{\Theta} U_A ( a, \theta ) P_A(\theta | a, d) \dd \theta,
     \end{aligned}
     \end{equation*}
     where $\Psi(a,d)$ represents the random expected utility associated with the couple $(a,d)$,
      for the random utility $U_A$ and the random probability $P_A$.
Finally, the defender finds $p(a\vert d)$, as the density associated with the distribution function $ \mathbb{P}_F ( A^*(d)\leq a)$ and uses it to solve (\ref{kantorei}).
 
Despite its applied relevance, sketched in Section 1, including reliability and risk analysis settings with adversaries
affecting systems safety and security, the aforementioned ARA approach faces two problems when trying to estimate $p_{D}(a \vert d)$:
 \begin{enumerate}[label=\alph*)]
     \item Modelling the random utility $U_A(a,\theta)$ entails prior knowledge and information about the attacker's preferences, which may be hard to elicit.
     \item Solving for problem \begin{small}    
     $\argmax _{a \in \mathcal{A}}\Psi_A(a,d)$\end{small} requires complex simulation, entailing repeatedly sampling from the 
     attacker's random utility and probability  and computing the optimal solution for each of these samples.
 \end{enumerate}

\subsection{ARA-FPD in sequential defend-attack games}\label{saturday}
We mitigate both problems exploring how to estimate $p_D (a | d)$ drawing on FPD concepts. As a consequence we introduce 
a novel ARA-FPD scheme. FPD \citep{KARNYFPD2006} conceives the attacker as trying to influence the (conditional) density $p_A (\theta | a, d)$ of the outcome $\theta $, so that it approaches his ideal (desired) results modelled through a density $p^i_A (\theta | a, d)$.  The attacker would then select the attack $a^*$ that makes the (actual) density $p_{A}(\theta \vert a,d)$ as close as possible to such ideal density. 
   Closeness is quantified through the Kullback-Leibler \citeyearpar{KLDivOriginal} (KL) divergence, see \citet{KARNYFPDaxiomatization} for a justification. Thus, his best attack is assumed to be
\begin{equation}
    \begin{aligned}\label{eq:praga}
 a^* (d) = \argmin_{ a\in {\cal A}} KL [ p_A(\theta | a, d)|| p^i_A(\theta | a, d)]. 
   \end{aligned}
\end{equation}
This idea serves to mitigate ARA issue (a) above, as, in principle, it is simpler to obtain the ideal than the specific utility function.

As in standard ARA, the defender does not fully know the ideal density \begin{small}    
$p^i_A(\theta \vert a,d)$\end{small}, neither the actual density \begin{footnotesize}
$p_A(\theta \vert a,d)$\end{footnotesize} that the attacker employs. She models such uncertainty through random densities \begin{small} 
$F= (P^i_A(\theta \vert a,d),P_A(\theta \vert a,d))$\end{small}. This further mitigates ARA issue (a), as expressing uncertainty through a target density is more intuitive than explicitly defining the parameters of a specific complex random utility function $U_A$. 
Within $F$, $P_A(\theta \vert a,d)$ is somewhat easier to elicit from the defender: since both agents share the outcome space, one may assume the attacker's outcome model to be similar to that of the defender, suggesting that $P_A(\theta | a,d) $ could be derived by placing some uncertainty around its own $p_D(\theta | a,d)$. For example, if this distribution had a parametric form, we could place a distribution over the defender's own parameters. Alternatively, we could use a Dirichlet process \citep{DirichletProcess} with base measure $p_D(\theta | a,d)$ and a suitable concentration parameter. As for $P^i_A(\theta \vert a,d)$, the defender may have some prior information and knowledge about the attacker's preferred outcomes upon which she would build the corresponding distribution.  Then, to estimate $p(a\vert d)$, we would use Monte Carlo (MC) simulation, drawing $L$ samples $((p_A^i)^l,p_A^l) \sim F$;  find the optimal attacks
\begin{equation*}
    a^*_l(d) = \argmin_{a \in \mathcal{A}} \int_{\Theta} p^l_A(\theta|a,d) \log \left( \frac{p^l_A(\theta|a,d)}{p_A^{i\, l}(\theta \vert d)} \right) \dd\theta,
    \quad \forall d \in \mathcal{D};
\end{equation*}
\noindent and approximate $\hat{p}_D (a | d)$ as the density of the empirical distribution function 
\begin{equation*}
    \hat{F }_D(a \vert d) = \frac{\# \{ a_l^*(d) \leq a\}}{L},
\end{equation*}
which would then be fed into problem (\ref{kantorei}) to select the optimal defence. 

Let us illustrate the ideas with a simple yet realistic scenario.\footnote{The examples considered in the paper are discrete.
  We just replace
integrals and densities by sums and probability mass functions.}

\paragraph{Example 1.} 
Suppose  $D$ wants to choose a  weapon detection system in an airport. It has three options $\mathcal{D} = \{d_1,d_2,d_3 \}$, where $d_1$ represents an expensive but very effective weapon detector; $d_2$ is a cheaper yet not that effective detector; and, finally, $d_3$ represents installing no detector.
Similarly, assume there is an attacker $A$ with three options $\mathcal{A} = \{a_1,a_2,a_3 \}$, where $a_1$ represents choosing top-tier, expensive, harder-to-detect weapons; $a_2$, cheaper, but mid-tier easier-to-detect weapons; and $a_3$ means not bringing any weapons. As for the outcomes, consider $\Theta = \{\theta_1,\theta_2 \}$, where $\theta_1$ represents a failed attack and $\theta_2$, a successful one.  Table \ref{tab:DData} reflects the defender's probabilities for outcome $\theta_1$  (from which complementary probabilities for the other outcome are deduced) as well as the utilities for the defender.
\begin{table}[pos = h]
\centering
\caption{$D$'s probabilities and utilities in airport protection example.}
\label{tab:DData}
\begin{subtable}[t]{0.3\columnwidth}
\resizebox{\columnwidth}{!}{
\begin{tabular}{c|ccc}
$p_D(\theta_1\vert a_i,d_j)$ & $a_1$        & $a_2$        & $a_3$    \\ \hline
$d_1$    & 0.6  & 0.8 & 1 \\
$d_2$    & 0.2 & 0.6 & 1 \\
$d_3$    & 0      & 0      & 1 
\end{tabular}
}
\caption{$D$'s outcome probability}\label{tab:outcomeProbD}
\end{subtable}
\hspace{1cm}
\begin{subtable}[t]{0.18\columnwidth}
\resizebox{\columnwidth}{!}{
\begin{tabular}{c|cc}
$u_D$ & $\theta_1$ & $\theta_2$ \\ \hline
$d_1$   & -8                   & -12                  \\
$d_2$   & -4                    & -12                  \\
$d_3$   & 0                     & -12 
\end{tabular}
}
\caption{$D$'s utility}\label{tab:UtilityD}
\end{subtable}
\end{table}

 \noindent The random probabilities $P_A (\theta_1  | a,d)$ are modelled as beta distribution   $\beta e (\alpha _{a,d}, \beta _{a,d})$ with parameters $\alpha_{a,d}$ and $\beta _{a,d}$ chosen to match $p_D (\theta_1 | a,d)$ in Table \ref{tab:DData} as mean and with variance 0.01, and $P_A (\theta_2  | a,d)= 1 - P_A (\theta_1  | a,d)$. 
 
Assigning uncertainty on $A$'s ideal densities $P_A^i (\theta  | a,d)$  is harder and will require previous information, since it directly depends on $A$'s unknown intentions. For instance, $A$ could be believed to seek to cause the most harm, or, alternatively, just try to force $D$ to spend the most resources, or a mixture of both. 
For simplicity, consider only one ideal density in which the attack always succeeds. It is associated with $A$ trying to do the most harm, that is
 $p^i_A(\theta_1 \vert a_i,d_j) = 0, \hspace{5pt} p^i_A(\theta_2 \vert a_i,d_j) = 1, \hspace{5pt} \forall i,j = 1,2,3.$
However, since KL divergences are not finite when $p^i_A=0$ and $p_A \neq 0$,
we soften $p^i_A$ through Laplace smoothing \citep{smoothing} making
$ p^i_{sA}(\theta\vert a,d) = \frac{p^i_A(\theta\vert a,d) + s}{1 + s \cdot \#(\Theta)},$%
with smoothing parameter $s = 0.01$ (and outcome space cardinality $\#(\Theta)=2$).

 Under these assumptions, the defender can model the attacker's action selection for each defence $d_j \in \mathcal{D}$, accounting for the uncertainty in $P_A$, by sampling $L$ times from  $\{p^l_A (\theta | a,d_j)\}_{l=1}^L\sim \beta e (\alpha _{a,d_j},\beta _{a,d_j})$ for all attacks $a\in \mathcal{A}$, calculating for each sample the attack minimising KL as in (\ref{eq:praga}) through
 \begin{equation*}
a^*_l(d_j) = \argmin_{a \in \mathcal{A}} \sum_{\theta \in \Theta} p^l_A(\theta|d_j,a) \log \left( \frac{p^l_A(\theta|d_j,a)}{p_{sA}^i(\theta \vert d_j,a)} \right),\ \forall d_{j}\in \mathcal{D},
\end{equation*}
\noindent and, with that, estimating $p(a\vert d_j)$ through the empirical frequencies. Table \ref{tab:pAestimate} shows the estimated probabilities for a simulation with sample size $L=10000$, portraying an aggressive attacker who tries to cause the most harm, as idealised.
\begin{table}[pos =h]
\centering
\caption{$D$'s estimate on $A$'s attack response probability}
\label{tab:pAestimate}
\resizebox{0.3\columnwidth}{!}{
\begin{tabular}{c|ccc}
$\hat{p}_D(a_i\vert d_j)$ & $a_1$        & $a_2$        & $a_3$    \\ \hline
$d_1$    & 0.84  & 0.16 & 0 \\
$d_2$    & 0.96 & 0.04 & 0 \\
$d_3$    & 0.5      & 0.5      & 0
\end{tabular}
}
\end{table}

 \noindent $D$ incorporates the estimate $\hat{p}_D (a\vert d)$ 
  to calculate her optimal defence in  (\ref{kantorei}), expressed in the discrete case 
    through 
    \begin{equation*}
    d^* = \argmax_{d\in \mathcal{D}} \psi_D (d) = \sum_{\theta\in \Theta}\sum_{a\in \mathcal{A}} u_D(d,\theta)p_D(\theta\vert a,d)\hat{p}_D (a\vert d).\end{equation*}
  The expected utilities of the three defences are, respectively, $\psi_D(d_1) = -9.47$, $\psi_D(d_2) = -10.28$, $\psi_D(d_3) = -12$.
Therefore, the optimal defence is
    $d_1 = \argmax_{d\in \mathcal{D}} \hspace{0.1cm}\psi_D(d).$ \hfill $\triangle$

\subsection{A complementary ARA-FPD approach } \label{sec:State_action_controlSeq}
 
 This section presents a complementary ARA-FPD approach.
 In our Section \ref{saturday} formulation, we used preferences based only on the outcomes $\theta$.
  That caused that, e.g., in Example 1 $a_1$ and $a_2$ were not actually distinguishable in preference terms, even though $a_2$ was cheaper.
 In turn, the formulation in this section allows the defender to better assess the attacker's problem. However, for it to be effective, $D$ must have additional prior knowledge about $A$, which may not always be the case. Both formulations are thus useful, and choosing one over the other will essentially depend on the problem context.
 
 Let us therefore reformulate the attacker's problem to simultaneously control the outcome $\theta \in \Theta$ and 
  place additional preferences on the attacker's action $a \in \mathcal{A}$.
Adapting the idea from Section \ref{saturday}, $A$ aims for a density $p_A(\theta,a \vert d)$ minimising  (compare with (2))

\begin{equation} \label{eq:state_act_control_seq}
    \begin{aligned}
   &KL\left[ p_A(\theta,a|d)\|p^i_A(\theta,a|d)\right]= KL \left[ p_A(\theta\vert a,d) p_A(a \vert d)\|p^i_A(\theta \vert a,d)p^i_A(a\vert d)\right],
   \end{aligned}
\end{equation}
where we have factorised $p_A(\theta,a \vert d)$ into the attacker's assessment of the outcome density 
$p_A(\theta \vert a,d)$ and his response to action $d$, designated $p_A(a \vert d)$, and have similarly decomposed the ideal outcome density. 
Elaborating expression (\ref{eq:state_act_control_seq}) we get
\begin{small}
\begin{equation}  \label{eq:DKLprev}
    \begin{aligned}
       & KL [p_A(\theta\vert a,d) p_A(a \vert d)\|p^i_A(\theta \vert a,d)p^i_A(a\vert d)] = 
          \int_{\mathcal{A}} \int_{\Theta} p_A(\theta|a,d) p_A(a|d) \left( \log \left( \frac{p_A(a|d)}{p_A^i(a|d)} \right)+ \log \left( \frac{p_A(\theta|a,d)}{p_A^i(\theta|a,d)} \right) \right)\dd \theta \dd a \\
          &= \left [\int_{\mathcal{A}} p_A(a|d) \log \left( \frac{p_A(a|d)}{p_A^i(a|d)} \right) \underbrace{\int_{\Theta} p_A(\theta \vert a,d)\dd \theta}_{=1} \dd a \right ] + \int_{\mathcal{A}} p_A(a|d) \int_{\Theta} p_A(\theta \vert a,d) \log \left( \frac{p_A(\theta|a,d)}{p_A^i(\theta|a,d)} \right) \dd \theta \dd a \\
           & = KL [ p_A(a \vert d) \| p^i_A (a \vert d)] + \int_{\mathcal{A}} p_A(a|d) KL[p_A(\theta \vert a,d) \| p^i_A(\theta \vert a,d)] \dd a .
    \end{aligned}
\end{equation}
\end{small}
\noindent Hence, for each defence $d\in \mathcal{D}$,  
  we just need to find the attack density $\bar{p} \in \Gamma \subseteq \mathcal{P}(\mathcal{A})$ minimising expression (\ref{eq:DKLprev}), that is
\begin{small}
\begin{equation} \label{eq:VI_Dkl}
\begin{aligned}
    \argmin_{\bar {p} \in \Gamma } \Bigg[& KL [\bar {p} (a|d) \| p^i_A (a \vert d)]  + \int_{\mathcal{A}} \bar {p}
     (a|d) 
    KL [p_A(\theta \vert a,d) \| p^i_A(\theta \vert a,d)] \dd a\Bigg],
    \end{aligned}
\end{equation}
\end{small}
where    $\mathcal{P}(\mathcal{A})$ represents the set of densities over ${\cal A}$  and $\Gamma$ is the subset where 
the search  would be constrained, e.g.,  a finite  
set of densities or a variational family (such as a family of Gaussians) as in variational inference \citep{blei2017variational}.

Intuitively, problem (\ref{eq:VI_Dkl}) can be understood as the attacker trying to select his attack as close as possible to his preferred choice, while also controlling his attack selection to achieve (on average) outcome(s) as close as possible to his ideal. 
Interestingly, such problem has an explicit solution. We include a proof of it,
 adapted from earlier related results in \citet{KARNYFPD2006}, \citet{vsindelavr2008stochastic}, and \citet{VIGeneralisation}.

\begin{prop} The solution of (\ref{eq:VI_Dkl}) is
\begin{small}
\begin{equation} \label{KL_DivExplicitsol}
\begin{aligned}
    &p_A (a \vert d) = \frac{p^i_A(a \vert d)  \exp\{-KL[p_A(\theta \vert a,d) \| p_A^i(\theta \vert a, d)]\}}{Z} \propto  p^i_A(a \vert d) \exp\{-KL[p_A(\theta \vert a,d) \| p_A^i(\theta \vert a, d)]\}
\end{aligned}
\end{equation}
\end{small}
 with $Z = \int _{\mathcal{A}} p^i_A(a)  \exp\{-KL[p_A(\theta \vert a,d) \| p_A^i(\theta \vert a, d)]\}\dd a$, assuming $Z<\infty$.
\end{prop}
\vspace{0.2in}

\noindent Importantly, this greatly alleviates ARA issue (b) from Section 2.1. While standard ARA requires solving a potentially expensive numerical optimization problem ($\argmax_{a \in \mathcal{A}}\Psi_A(a,d)$) for every $d$ for every simulated sample, the FPD formulation bypasses this step by using the explicit minimiser (\ref{KL_DivExplicitsol}) to estimate $p_A (a\vert d)$.\footnote{Importantly, especially in continuous problems, for optimization purposes in (1), we do not actually need to compute the expensive $Z$.}  
 As in Section \ref{saturday}, the defender places uncertainty on $A$'s ideal outcome $p_A^{i\theta}(\theta \vert a,d)$ and action $p^{ia}_A(a \vert d)$
  densities, as well as on $p_A(\theta \vert a,d)$, through a distribution over densities $F=(P^{i\theta}_A(\theta \vert a,d),P^{ia}_A(a\vert d),P_A(\theta \vert a,d)).$ 
Then, the defender uses MC simulation, drawing $L$ samples $((p^{i\theta}_A,p^{ia}_A,p_A)^l)$ 
from $F$, finding\footnote{For notational simplicity, and whenever the meaning is clear from context, we shall write $p^i_A$
 in place of both $p^{i\theta}_A$ and $p^{ia}_A$.} 
\begin{equation} \label{eq:att_estimate_state_action_seq}
p^*_l(a \vert d) \propto p^{il}_A (a \vert d) \exp\{KL [p^l_A(\theta \vert a,d) \| p^{il}_A(\theta \vert a, d)]\},\; \forall d\in \mathcal{D} .
\end{equation}
  With that, the estimate 
\begin{equation} \label{eq:att_estimate_state_actionMC_seq}
    \hat{p}_A(a \vert d) = \frac{\sum_{l= 1}^Lp^*_l(a\vert d)}{L}
    \end{equation}
is finally injected into (\ref{kantorei}) to select the defender's optimal decision.

We illustrate the complementary formulation through the following 
 example.
\paragraph{Example 2.} 
Besides the data from Example 1, we also have to care about the attacker's ideal action-selection probability, $p^i(a\vert d)$. It provides prior knowledge about the attacker's intentions. For comparison, consider two cases. First, if the attacker aims to maximize his attack success, while not being caught, we define his idealized probability as in Table \ref{tab:pAiMaxAtt}. Second, should he aim to force the defender to spend the most with him spending the least, his ideal probabilities will be defined as in Table \ref{tab:pAiMaxDefCost}. 

\begin{table}[pos = h]
\centering
\caption{Two examples for $A$'s ideal action selection probabilities.}
\label{tab:pAi}
\begin{subtable}[t]{0.3\columnwidth}
\resizebox{\columnwidth}{!}{
\begin{tabular}{c|ccc}
$p_A^i( a_i \vert d_j)$ & $a_1$        & $a_2$        & $a_3$    \\ \hline
$d_1$    & 0.8  & 0.2 & 0 \\
$d_2$    & 0.7 & 0.3 & 0 \\
$d_3$    & 0      & 1      & 0 
\end{tabular}
}
\caption{}
\label{tab:pAiMaxAtt}
\end{subtable}
\hspace{0.5cm}
\begin{subtable}[t]{0.3\columnwidth}
\resizebox{\columnwidth}{!}{
\begin{tabular}{c|ccc}
$p_A^i( a_i \vert d_j)$ & $a_1$        & $a_2$        & $a_3$    \\ \hline
$d_1$    & 0.2  & 0.1 & 0.7 \\
$d_2$    & 0.6 & 0.1 & 0.3 \\
$d_3$    & 0      & 1      & 0 
\end{tabular}
}
\caption{}
\label{tab:pAiMaxDefCost}
\end{subtable}
\end{table}

\noindent  Under these assumptions, the defender models the attacker's action selection for a given defence $d_j \in \mathcal{D}$, accounting for uncertainty on $p_A (\theta | a, d_j)$ by sampling $L$ times from  $\{p^l_A (\theta | a,d_j)\}_{l=1}^L\sim \beta e (\alpha _{a,d_j},\beta _{a,d_j})$ for all $a\in \mathcal{A}$, with $\alpha _{a,d_j}$ and $\beta _{a,d_j}$ from Example 1, calculating for each sample the probabilities through (\ref{eq:att_estimate_state_action_seq}).
   With that, we estimate $p_A(a\vert d_j)$ through (\ref{eq:att_estimate_state_actionMC_seq}).

Table \ref{tab:pA} shows the estimated probabilities for a simulation with $L=10000$ in both scenarios. Observe how Table \ref{tab:pAMaxAtt}, corresponding to Table \ref{tab:pAiMaxAtt}, suggests an aggressive attacker, whereas Table \ref{tab:pAMaxDefCost}, cf.\ Table \ref{tab:pAiMaxDefCost}, reflects a more conservative attacker, deterred when $D$ chooses $d_1$. Furthermore, observe how placing the attack ideal prior leads the attacker to select $a_2$ when the defender chooses $d_3$, unlike in Example 1.
\begin{table}[pos = h]
\centering
\caption{Defender's estimates on $A$'s responses}
\label{tab:pA}
\begin{subtable}[t]{0.3\columnwidth}
\resizebox{\columnwidth}{!}{
\begin{tabular}{c|ccc}
$\hat{p}_A(a_i\vert d_j)$ & $a_1$        & $a_2$        & $a_3$    \\ \hline
$d_1$    & 0.89  & 0.11 & 0 \\
$d_2$    & 0.9 & 0.1 & 0 \\
$d_3$    & 0      & 1      & 0
\end{tabular}
}
\caption{}
\label{tab:pAMaxAtt}
\end{subtable}
\hspace{0.5cm}
\begin{subtable}[t]{0.3\columnwidth}
\resizebox{\columnwidth}{!}{
\begin{tabular}{c|ccc}
$\hat{p}_A(a_i\vert d_j)$ & $a_1$        & $a_2$        & $a_3$    \\ \hline
$d_1$    & 0.63  & 0.14 & 0.22 \\
$d_2$    & 0.95 & 0.04 & 0.01 \\
$d_3$    & 0      & 1      & 0
\end{tabular}
}
\caption{}
\label{tab:pAMaxDefCost}
\end{subtable}
\end{table}

\noindent Finally, $D$ incorporates these estimates to calculate her optimal defence through problem (\ref{kantorei}). The defences' expected utilities obtained for the probabilities in Table \ref{tab:pAMaxAtt} are $\psi_D(d_1) = -9.5$, $\psi_D(d_2) = -10.08$, $\psi_D(d_3) = -12$, and, hence, the optimal defence is $d_1$. Those
 for the probabilities in Table \ref{tab:pAMaxDefCost} are $\psi_D(d_1) = -9.13$, $\psi_D(d_2) = -10.2$, $\psi_D(d_3) = -12$, again being $d_1$ the optimal defence.
\hfill $\triangle$
\vspace{0.2in}

\noindent Observe that despite providing 
  a more detailed estimate of $A$'s attack probability, 
   the optimal defence is still $d_1$. The following section studies a case where the optimal defence 
     differs and further compares the results obtained.
\subsection{ARA-FPD in simultaneous defend-attack games} \label{sec:SimultaneousARAFPDSingle}
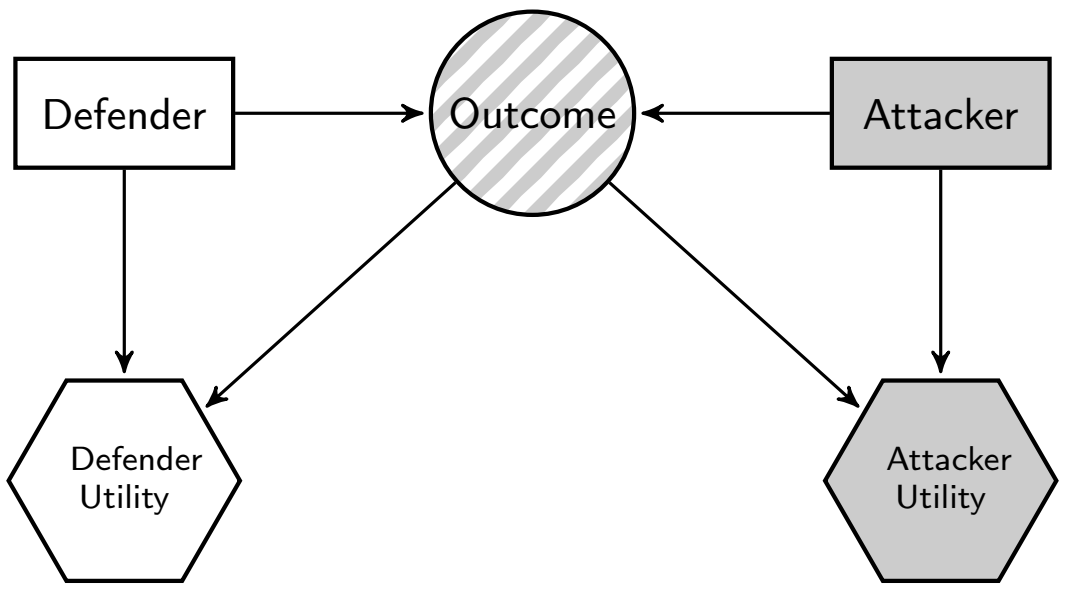
\begin{figure}[pos = htbp]
\centering
\begin{tikzpicture}[->,>=stealth',shorten >=1pt,auto,node distance=2.5cm, scale=0.9, transform shape, semithick]

  \tikzstyle{decisionA} = [rectangle, thick, minimum width=1.6cm, minimum height=0.8cm, draw=black, fill=gray!40, align=center]
  \tikzstyle{decisionD} = [rectangle, thick, minimum width=1.6cm, minimum height=0.8cm, draw=black, align=center]

  \tikzstyle{utilityA} = [regular polygon, regular polygon sides=6, thick, minimum size=1.7cm, draw=black, fill=gray!40, align=center, inner sep=1pt, text width=0.8cm, font=\scriptsize]
  \tikzstyle{utilityD} = [regular polygon, regular polygon sides=6, thick, minimum size=1.7cm, draw=black, align=center, inner sep=1pt, text width=0.8cm, font=\scriptsize]

  \tikzstyle{chance} = [circle, thick, minimum size=0.6cm, inner sep=1pt, draw=black, pattern=stripes, pattern color=gray!40, align=center, text width=1.4cm, font=\small]

  \node[chance](D) {Outcome};
  \node[decisionD] (A) [left of=D, xshift=-0.5cm] {Defender};
  \node[decisionA] (B) [right of=D, xshift=0.5cm] {Attacker};
  \node[utilityD]  (C) [below of=A, yshift=-0.2cm] {\centering Defender \\ Utility};
  \node[utilityA]  (E) [below of=B, yshift=-0.2cm] {\centering Attacker \\ Utility};

  \path (A) edge    node {} (D)
            edge    node {} (C)
        (B) edge    node {} (D)
            edge    node {} (E)
        (D) edge    node {} (C)
            edge    node {} (E);
\end{tikzpicture}
\caption{Basic two-player simultaneous defend-attack game}
\label{fig:sim-Game}
\end{figure}

Consider now simultaneous defend-attack interactions. As
  in Figure~\ref{fig:sim-Game}, both agents 
 decide without knowing the action chosen by each other,  reflected through the lack of an arc from the Defender to the Attacker node.

\noindent An ARA formulation for this game supports $D$'s decisions via maximizing her expected utility through
\begin{equation}\label{kantorei2}
d^* =  \operatorname*{arg\,max}_{d \in {\cal D}} \,
       \int_{\mathcal{A}}  \psi_D (d, a) p_D (  a ) \dd a,
\end{equation}
where $p_D(a)$ models her beliefs about the attacker's decision $a$ and $\psi_D(d,a)$ is as in (\ref{kantorei}). 
As in Section \ref{saturday}, in its basic form, ARA assumes that the defender believes the attacker maximizes expected utility. However, in this type of interaction, the attacker does not know $D$'s defence when selecting his attack $a$. Hence, he is believed to place a density $p_A(d)$
 on $D$'s decision $d$, and choose his attack according to a formulation symmetric to (\ref{kantorei2})
 \begin{equation*}    
 a^* = \argmax_{a \in {\cal A}} 
     \int_{\mathcal{D}}   \left[ \int_{\Theta}  u_A ( a, \theta ) 
 p_A (\theta  |  d,a) \dd \theta \right]  p_A ( d ) 
      \dd d. \end{equation*}
 
\noindent In general, $D$ has incomplete information about $A$'s required inputs $(u_A (a,\theta),$ $ p_A (\theta |d,a), p_A (d))$. 
Suppose she models all the information available to her about these elements in a Bayesian manner through a distribution $F=(U_A (a,\theta),P_A (\theta |d,a), \Pi_A (d))$ over random utilities and probabilities,  where $\Pi_A (d)$ corresponds to the random version of 
 $p_A(d)$. This uncertainty is propagated to the distribution of optimal attacks
 \begin{small}
\begin{equation*}
    A ^* \sim  \argmax_{a \in {\cal A}} \,
    \int_{\mathcal{D}}   \left[ 
    \int_{\Theta}  U_A ( a, \theta ) \;
    P_A (\theta  \mid d,a)  \dd \theta  \right]
  \,    \Pi_A ( D = d ) \dd d.
\end{equation*}
\end{small}
While $P_A(\theta |d,a)$ is simpler to elicit due to its structural similarity with $p_D(\theta  \vert a,d)$, $U_A(a,\theta)$ must rely on prior knowledge and information not always available. Furthermore, eliciting $\Pi_A(d)$ may require a deeper analysis as 
 the defender needs to model how the attacker analyses $D$'s problem: this potentially leads to a recursion of nested decision models, significantly increasing computational needs. However, we shall limit the analysis to just one level of the recursion making the defender a level-$2$ thinker in the sense of \cite{stahl1994experimental}.
 
 Again, to mitigate the problems about modelling $U_A(a,\theta)$, we employ an ARA-FPD approach, similar in spirit to that in
 Section \ref{saturday}.
   The attacker sets his preferences wrt the outcome $\theta $. In this case, since the attacker does not observe the defender's action $d$, it is assimilated to choosing his attack according to  
   \begin{equation} \label{eq:sim_single_stage_outcome} 
   \argmin_{a \in \mathcal{A}} KL [p_A(\theta,d \vert a)\|p^i_A(\theta,d \vert a)].
   \end{equation}
Since the attacker lacks specific information or strict preferences concerning the defender's strategy, we assume that their ideal distribution for $d$ matches their actual distribution, i.e. $ p^i_A(d) = p_A(d)$. Within the FPD jargon, equating an ideal density with its actual density is referred to as {\em leaving the variable 'to its fate'}. This simplifies the attacker's decision in (\ref{eq:sim_single_stage_outcome}) to
\begin{equation*}    
\argmin_{a \in \mathcal{A}} \int_{\mathcal{D}} KL [p_A(\theta \vert d,a)\|p^i_A(\theta \vert d, a)] p_A(d) \dd d.\end{equation*}
Consequently, when the defender models the attacker's problem,  
  she has to place as well uncertainty on $A$'s modelling distribution over her own behaviour. Hence, the defender models the uncertainty through a random distribution 
  \begin{small}    
  $ G=  (P_A(\theta \vert a,d), $ $ P^{i}_A(\theta\vert a, d), \Pi_A(d)) $\end{small}. While \begin{small}    
  $(P_A(\theta \vert a,d),P^{i}_A(\theta \vert a,d))$ \end{small} may be elicited as in the sequential defend-attack interaction (Section \ref{saturday}), $\Pi_A(d)$ potentially leads to a recursion of nested models. However recall that we only consider one level within this hierarchy.

 The estimation of $p_A(a)$ is undertaken through MC.
First, draw $K$ samples $\{\pi_A^k(d) \}_{k=1}^K \sim \Pi_A(d)$, and, from each, draw $N$ samples $\{d^k_n\}_{n=1}^N \sim \pi_A^k(d)$. Then, for every $d^k_n$, we sample $L$ times $\{ (p_A(\theta \vert a,d_n^k),p_A^i(\theta \vert a,d_n^k))^l\}_{l=1}^L \sim (P_A,P^i_A)$ and find, for each $l$,
\begin{equation*}
    a_{lkn}^* = \argmin _{a \in \mathcal{A}} KL \left[ (p^l_A(\theta|a,d^k_n)\|(p^i_A)^l(\theta|a,d^k_n)\right].\end{equation*}
 We then use the $\{ a_{lkn}^* \}$
 to approximate the distribution function associated with $p_D(a)$ using the empirical distribution function 
 $$\frac{\# \{ a_{lkn}^* \leq a\}}{L\cdot K \cdot N}.$$

\paragraph{Example 3.} 
Consider Example 1 from a simultaneous perspective, thus requiring $\Pi_A(d)$. For simplicity, suppose that the attacker samples $D$'s decisions uniformly. The remaining data and parameters are as in Example 1.
 Under these assumptions, $D$ can model $A$'s action selection: for each random sample $d^n\in \mathcal{D}$ (drawn $N$ times), she accounts for her uncertainty about $P_A(\theta \vert a,d)$ by sampling $L$ times from  $\{p^l_A (\theta | a,d^n)\}_{l=1}^L\sim \beta e (\alpha _{a,d^n},\beta _{a,d^n})$ for all $a\in \mathcal{A}$, calculating for each sample
$$a^*_{ln} = \argmin_{a \in \mathcal{A}} \sum_{\theta } p^l_A(\theta|a,d^n) \log \left( \frac{p^l_A(\theta|a,d^n)}{p_{sA}^i(\theta \vert a,d^n)} \right). $$
 With that, $D$ estimates $p_A (a)$ via empirical frequencies. The estimated probabilities with $L=10000$ and $N$ = 100 ($K=1$ since we are not placing uncertainty on $D$'s defence selection for simplicity) are $\hat{p}_A (a_1) = 0.9$, $\hat{p}_A(a_2) = 0.08$, and $\hat {p}_A (a_3) = 0.02$. Hence, the expected utilities are $\psi(d_1) = -9.5$, $\psi(d_2) = -10.01$, and $\psi(d_3) = -11.76$, and
  the optimal defence is $d^* = \argmax_{d\in \mathcal{D}} \hspace{0.1cm}\psi(d) = d_1.$ \hfill $\triangle$
\vspace{0.2in}

\noindent Similar to Section \ref{sec:State_action_controlSeq}, the aforementioned approach can be reformulated as an outcome-action assessment problem. Once again, since the attacker does not observe the defender's action $d$ before acting, he evaluates his strategy through the joint distribution
\begin{equation} \label{eq:sim_single_stageKL}
\argmin_{p_A(a) \in \Gamma}KL[p_A(a,\theta,d)\, \| \, p^i_A(a,\theta,d)].
\end{equation}
 As before, assuming that the attacker has no strict preferences over the defender's strategy, as introduced earlier, we \textit{leave $p_A(d)$ to its fate}, that is, we assume $p_A^i(d) = p_A(d)$, and the divergence in (\ref{eq:sim_single_stageKL}) expands to

\begin{small}
\begin{equation} \label{eq:single_stage_simultaneous_prev}
    \begin{aligned}
       &KL[p_A(a,\theta,d) \| p^i_A(a,\theta,d)]= \\& \left [\int_{\mathcal{A}}\int_{\mathcal{D}} p_A(d) p_A(a|d) \log \left( \frac{p_A(a|d)}{p_A^i(a|d)} \right) \underbrace{\int_{\Theta} p_A(\theta \vert a,d)\dd \theta}_{=1} \dd a \dd d\right ] + \int_{\mathcal{A}}\int_{\mathcal{D}} p_A(a|d) p_A(d)\int_{\Theta} p_A(\theta \vert a,d) \log \left( \frac{p_A(\theta|a,d)}{p_A^i(\theta|a,d)} \right) \dd \theta \dd a \dd d,
    \end{aligned}
\end{equation}
\end{small}
derived as in (\ref{eq:DKLprev}).
Furthermore, the simultaneous nature of the interaction dictates that the attacker's action selection cannot be influenced by the defender's action. This structural independence implies $p_A(a\vert d) = p_A(a)$ and $p_A^i(a\vert d) = p_A^i(a)$. Applying these simplifications to (\ref{eq:single_stage_simultaneous_prev}) reduces the objective to
\begin{small}
\begin{equation} \label{eq:single_stageKL}
    KL [p_A(a) \| p^i_A (a)] + \int_{\mathcal{A}} p_A(a) \mathbb{E}_d\left[KL [p_A(\theta \vert a,d) \| p^i_A(\theta \vert a,d)]\right] \dd a.
\end{equation}
\end{small}
Analogous to Proposition 1, problem (\ref{eq:single_stageKL}) has the explicit solution
\begin{equation*}
    \begin{aligned}
    &p^*_A (a) = \frac{p^i_A(a)  \exp\{-\mathbb{E}_d \left[KL[p_A(\theta \vert a,d) \| p_A^i(\theta \vert a, d)]\right]\}}{Z} \propto p^i_A(a) \exp\{-\mathbb{E}_d\left[KL[p_A(\theta \vert a,d) \| p_A^i(\theta \vert a, d)]\right]\},
    \end{aligned}
\end{equation*}
which can be used to estimate $p_A(a)$.
The defender handles her uncertainty over $A$'s decision through a random distribution
$F=(P_A(\theta \vert a,d),P^{i}_A(\theta \vert a,d),P^{i}_A(a),\Pi_A(d))$. 
Finally, MC serves to estimate $p_D(a)$:  
we draw $K$ samples $\{\pi_A^k(d) \}_{k=1}^K \sim \Pi_A(d)$, and $N$ samples $\{d^k_n\}_{n=1}^N \sim \pi_A^k(d)$.  Then, for every $d^k_n$, we sample $L$ times $$\{ ((p^{i\theta}_A(\theta \vert a,d_n^k),p^{ia}_A(a),p_A(\theta \vert a,d_n^k))^l)\}_{l=1}^L \sim (P^{i\theta}_A,P^{ia}_A,P_A)$$ and find, for each $l$,
\begin{equation} \label{eq:MCMCSimultaneousStateactionControl}    
p^*_{lkn}(a) \propto p^{il}_A(a)  \exp\{-KL[(p^l_A(\theta \vert a,d^k_n]) \| p^{il}_A(\theta \vert a, d^k_n)]\},
\end{equation}
and estimate $\hat{p}_{A}(a) = \frac{\sum_{k=1}^K\sum_{l= 1}^L\sum_{n=1}^Np^*_{lkn}(a)}{L\cdot K \cdot N}.$

 Let us next compare various formulations.   

\paragraph{Example 4.}
  Consider the simultaneous setting of Example 2. Hence ideals $p^i(a)$ are not conditioned on $d$. Consider two scenarios, $p_A^i(a) = (0.7,0.3,0)$ and $p_A^i(a) = (0.3,0.7,0)$, with the remaining parameters as in Example 2.
 $D$ models the attacker's action selection through sampling as in Example 3,
 estimating $p^*_{ln}(a)$ through (\ref{eq:MCMCSimultaneousStateactionControl}) (with $K=1$ since we are not placing uncertainty on $D$'s defence selection). 

With that, we estimate $p_A(a)$.
The estimated probabilities for $L=10000$ and $N$ = 100 are $\hat{p}_A(a_1) = 0.82$, $\hat{p}_A(a_2) = 0.18$ and $\hat{p}_A(a_3) = 0$ for the former idealized actions, and $\hat{p}_A(a_1) = 0.52$, $\hat{p}_A(a_2) = 0.48$ and $\hat{p}_A(a_3) = 0$ for the latter. These probabilities lead, respectively, to the expected utilities $\psi(d_1) = -9.4$, $\psi(d_2) = -9.9$, $\psi(d_3) = -12$ and $\psi(d_1) = -9.2$, $\psi(d_2) = -8.8$, $\psi(d_3) = -12$. 
 Then, the optimal defence in the first case is $d_1$ and, in the second, is $d_2.$
\hfill $\triangle$
\vspace{0.2in}

\noindent 
Interestingly, Example 3 still shows a very aggressive attacker, which makes sense since we are only focusing on the idealized goal of always winning, leading to a strong defence, $d_1$. In turn, Example 4 suggests more conservative attackers, leading to 
  defence $d_2$. 

It is useful to compare the optimal defences across the four examples. In the sequential setting (Examples 1 and 2), the attacker always knows $D$’s choice, which drives the defender towards the more conservative option $d_1$, independent of the formulation. In contrast, in simultaneous cases (Examples 3 and 4), the joint action-outcome control introduces dependencies that refine the attacker’s modelled behaviour and, in turn, the defender’s response (e.g., $d_2$ in Example 4). This suggests that richer formulations can alter defensive recommendations by capturing subtleties missed in outcome-only models.
\subsection{Connecting ARA and ARA-FPD} \label{sec:Connecting_ARA_FPD}

While the conventional ARA framework relies on the explicit elicitation and maximization of expected utilities, the ARA-FPD methodology reformulates the attacker's problem as minimizing the Kullback-Leibler (KL) divergence to an ideal probability distribution. By interpreting the logarithm of the attacker's ideal outcome distribution as a utility function, that is,
\begin{equation} \label{eq:ImplicitUtilityOutcome}
    u_A(a, \theta, d) = \log \left( \frac{p^i_A(\theta \vert a, d)}{p_A(\theta \vert a, d)} \right),
\end{equation}
the FPD framework can be related to the ARA expected utility structure. This connection reveals that the  FPD formulation in Section \ref{saturday} (outcome based) is equivalent to a standard ARA with a specific utility function, whereas the one in Section \ref{sec:State_action_controlSeq} (outcome-action) is its soft version, i.e., it operates as an entropy-regularized expected utility maximiser. The following propositions formalise these connections for both sequential and simultaneous interactions. Let us start with the outcome based setup

\begin{prop} \label{thm:HardEquivalence}
Under utility function (\ref{eq:ImplicitUtilityOutcome}), the outcome-based FPD formulations are equivalent to conventional ARA expected utility maximization problems, that is
\begin{enumerate}[label=\alph*)]
    \item Sequential: $\argmin_{a \in \mathcal{A}} KL[p_A(\theta \vert a, d) \| p^i_A(\theta \vert a, d)] \equiv \argmax_{a \in \mathcal{A}} \psi_A(d,a)$.
    \item Simultaneous: $\argmin_{a \in \mathcal{A}} \int_{\mathcal{D}} KL[p_A(\theta \vert a, d) \| p^i_A(\theta \vert a, d)] p_A(d) \dd d \equiv \argmax_{a \in \mathcal{A}} \psi_A(a)$.
\end{enumerate}
\end{prop}
\vspace{.1in}
Next, we now show that expanding the FPD control to include the attacker's action density modifies the optimization from a hard maximizer to a smoothed (soft) entropy-regularized maximizer.

\begin{prop} \label{thm:SoftEquivalence}
Consider the ARA-FPD approach (\ref{eq:state_act_control_seq} and \ref{eq:sim_single_stageKL}) defining a joint ideal distribution over actions and outcomes. Then, the outcome-action ARA-FPD optimization is equivalent to maximizing an entropy-regularized expected utility, that is

\begin{enumerate}[label=\alph*)]
    \item Sequential: The ARA-FPD problem $p^*_A = \argmin_{p_A \in \Gamma} KL\left[p_A(a, \theta \vert d) \| p^i_A(a, \theta \vert d)\right]$ is equivalent to
    \begin{equation} \label{eq:SoftSequential}
        p^*_A = \argmax_{p_A \in \Gamma} \int_{\mathcal{A}} p_A(a \vert d) \Big[ \psi_A(d,a) + \log p^i_A(a \vert d) \Big] \dd a + \mathcal{H}(p_A(a \vert d)).
    \end{equation}
    
    \item Simultaneous: Assuming $p^i_A(d) = p_A(d)$, the ARA-FPD problem $p^*_A = \argmin_{p_A \in \Gamma} KL\left[p_A(a, \theta, d) \| p^i_A(a, \theta, d)\right]$ is equivalent to
    \begin{equation} \label{eq:SoftSimultaneous}
        p^*_A = \argmax_{p_A \in \Gamma} \int_{\mathcal{A}} p_A(a) \Big[ \psi_A(a) + \log p^i_A(a) \Big] \dd a + \mathcal{H}(p_A(a)),
    \end{equation}
\end{enumerate}
where, the expected utilities $\psi_A(d,a)$ and $\psi_A(a)$ are as defined in Proposition \ref{thm:HardEquivalence}, with $\log p^i_A(\cdot)$ acting as a prior, and $\mathcal{H}(\cdot)$ denoting the Shannon entropy.
\end{prop}


These propositions show that extending ARA to ARA-FPD allows us to work with an entropy-regularized expected utility maximizer. Practically, this formulation serves to add a layer of smoothness to the predicted attack distribution while also mitigating the heavy burden of complex utility elicitation typical of ARA.

\subsection{A comparison} \label{sec:CaseStudy}

We end this section comparing the ARA-FPD methodology with a conventional ARA approach and a game-theoretic one. Consider a simultaneous defend-attack game framed around an air combat scenario where a defender $D$ manages an aerial defence system to protect a strategic infrastructure. 
The defender has three available security investments $\mathcal{D} = \{d_1, d_2, d_3\}$, where $d_1$ represents a substantial anti-aircraft missile deployment, $d_2$ is a moderate deployment, and $d_3$ is a minimal deployment. An adversary (attacker $A$) has three possible attack profiles $\mathcal{A} = \{a_1, a_2, a_3\}$, where $a_1$ represents a high-intensity drone strike, $a_2$ is a low-intensity one, and $a_3$ entails no attack. The outcome space $\Theta = \{\theta_1, \theta_2\}$ represents a failed attack ($\theta_1$) or a successful drone strike ($\theta_2$). 
$D$ assesses the probability of preventing the attack, $p_D(\theta_1 \vert a, d)$, and her utility $u_D(d, \theta)$ for each outcome, as Table \ref{tab:CaseStudyData} summarizes. 

\begin{table}[pos = htbp]
\centering
\caption{$D$'s outcome probabilities and utilities for the air combat scenario}
\label{tab:CaseStudyData}
\begin{subtable}[t]{0.45\columnwidth}
\centering
\begin{tabular}{c|ccc}
$p_D(\theta_1 \vert a, d)$ & $a_1$ & $a_2$ & $a_3$ \\ \hline
$d_1$ & $0.6$ & $0.8$ & $1.0$ \\
$d_2$ & $0.2$ & $0.6$ & $1.0$ \\
$d_3$ & $0.0$ & $0.0$ & $1.0$ \\
\end{tabular}
\caption{$D$'s outcome probability}
\end{subtable}
\hfill
\begin{subtable}[t]{0.45\columnwidth}
\centering
\begin{tabular}{c|cc}
$u_D(d, \theta)$ & $\theta_1$ & $\theta_2$ \\ \hline
$d_1$ & $-8.0$ & $-12.0$ \\
$d_2$ & $-4.0$ & $-12.0$ \\
$d_3$ & $0.0$ & $-12.0$ \\
\end{tabular}
\caption{$D$'s utility}
\end{subtable}
\end{table}

\noindent From these, $D$ computes her expected utility $\psi_D(d, a) = \sum_{\theta} u_D(d, \theta)p_D(\theta \vert a,d)$. Table \ref{tab:expUtMatrix} summarizes the expected utility matrix.
\begin{table}[pos = htbp]
\centering
\caption{Expected utility matrix}
\label{tab:expUtMatrix}
\centering
\begin{tabular}{c|ccc}
$\psi_D(d, a)$ & $a_1$ & $a_2$ & $a_3$ \\ \hline
$d_1$ & $-9.6$ & $-8.8$ & $-8.0$ \\
$d_2$ & $-10.4$ & $-7.2$ & $-4.0$ \\
$d_3$ & $-12.0$ & $-12.0$ & $-12.0$ \\
\end{tabular}
\end{table}

\textbf{Game Theory Solution}: Suppose we rely on a minimax game theoretic formulation, the defender of the problem being conservative against worst-case scenarios. Since $D$ lacks precise information regarding $A$'s operational preferences or budget, she operates under the assumption of common knowledge (CK) of a strictly zero-sum game. This means she assumes $A$'s objective is exactly to minimize her expected utility ($\psi_A(d,a) = -\psi_D(d,a)$). 

Observing the expected utility values $\psi_D(d, a)$, attack $a_1$ strictly dominates the attacker's choices in a zero-sum setting, as it guarantees the highest possible damage to the infrastructure regardless of the deployed defence. Consequently, anticipating this attack, $D$ selects her robust maximin defence $d^*$, which is the decision maximizing her utility against the worst-case attack
   $ d^* = \argmax_{d \in \mathcal{D}} \psi_D(d, a_1).$
Evaluating this across the available options yields $\psi_D(d_1, a_1) = -9.6$, $ \psi_D(d_2, a_1) = -10.4$, and $\psi_D(d_3, a_1) = -12.0$. Thus, the minimax formulation forces $D$ into the highly conservative strategy of selecting the substantial missile deployment $d_1$ reducing her expected losses to $\psi_D(d_1) = -9.6$.
\vspace{0.2cm}

\textbf{ARA Solution}: Conventional ARA drops the CK assumption, requiring $D$ to model $A$'s unknown beliefs and utilities. Since the interaction is simultaneous, $D$ accounts for $A$'s uncertainty about her deployed defences by placing a symmetric Dirichlet prior $\Pi_A(d) \sim \text{Dir}(10, 10, 10)$. Random probabilities $P_A (\theta_1 \vert a,d)$ are modelled as Beta distributions $\beta e (\alpha_{a,d}, \beta_{a,d})$ with parameters $\alpha_{a,d}$ and $\beta_{a,d}$ chosen to match $p_D (\theta_1 \vert a,d)$ in Table \ref{tab:CaseStudyData} as the mean and with variance 0.01, and $P_A (\theta_2 \vert a,d)= 1 - P_A (\theta_1 \vert a,d)$. Additionally, ARA requires $D$ to model $A$'s utility through a random utility function $U_A(a, \theta)$. Suppose $D$ decomposes this into a random reward for a successful drone strike $R$, and a random operational cost for executing the attack $C_a$, such that $U_A(a, \theta_2) = R - C_a$ and $U_A(a, \theta_1) = -C_a$.

Eliciting exact financial and operational costs for adversaries is highly complex, so $D$ is assumed to use broad uniform distributions. We explore two variants to illustrate the sensitivity of standard ARA to this utility elicitation:
\begin{enumerate}
    \item $D$ assesses $R \sim U(15, 25)$, $C_1 \sim U(4, 8)$ for the high-intensity strike $a_1$, $C_2 \sim U(1, 5)$ for the low-intensity strike $a_2$, and $C_3 = 0$. Through MC simulation, $D$ estimates $\hat{p}_A(a) = (0.67, 0.33, 0.0)$. Therefore, $D$ again selects $d_1$ ($\psi_D(d_1) = -9.34$).
    \item If $D$ alters her elicitation to assume $a_1$ is more expensive or logistically difficult for the attacker, shifting $C_1 \sim U(5, 10)$, the simulation returns $\hat{p}_A(a) = (0.42, 0.58, 0.0)$. Under this assumption $D$ finds optimal to switch to moderate deployment $d_2$ ($\psi_D(d_2) = -8.55$).
\end{enumerate}
\vspace{0.2cm}

\textbf{ARA-FPD Solution}: In this approach, $D$ avoids the elicitation of complex utility parameters. Instead, she places uncertainty over $A$'s ideal density $p^i_A(a)$, e.g.  using a Dirichlet distribution. Similar to previous examples, $D$ assumes $A$ ideally wants the drone strike to succeed ($p^i_A(\theta_2) = 1$, softened via Laplace smoothing).
Suppose the intelligence team suggests $A$ prefers the cheaper low-intensity drone strike ($a_2$), and therefore $D$ places uncertainty as $p^i_A(a) \sim \text{Dir}(30, 70, 10)$. Using the explicit minimizer and MC simulation, $D$ obtains an attack estimate of $\hat{p}_A(a) = (0.53, 0.47, 0.01)$. Evaluated against this estimate, using (\ref{kantorei2}), $D$ finds that the moderate deployment $d_2$ yields a better expected utility ($\psi_D(d_2) = -8.86$) than the substantial deployment $d_1$ ($\psi_D(d_1) = -9.21$). 
Conversely, if intelligence suggests $A$ has a preference for a high-intensity strike ($a_1$), modelled through a $\text{Dir}(70, 30, 10)$, the estimate shifts to $\hat{p}_A(a) = (0.82, 0.17, 0.0)$, making $D$'s decision to be $d_1$.
\vspace{0.2cm}

\textbf{Discussion}: Table \ref{tab:SummaryComparison} summarizes the attack probabilities and optimal aerial defences across the frameworks.

\begin{table}[pos = htbp]
\centering
\caption{Summary of expected utilities and optimal infrastructure defences}
\label{tab:SummaryComparison}
\resizebox{0.85\columnwidth}{!}{
\begin{tabular}{l|ccc|ccc|c}
\textbf{Method} & $p(a_1)$ & $p(a_2)$ & $p(a_3)$ & $\psi(d_1)$ & $\psi(d_2)$ & $\psi(d_3)$ & \textbf{Optimal} \\ \hline
Minimax (Game Theory) & $1.00$ & $0.00$ & $0.00$ & $\mathbf{-9.60}$ & $-10.40$ & $-12.00$ & $d_1$ \\
ARA ($C_1 \sim U(4,8)$) & $0.67$ & $0.33$ & $0.00$ & $\mathbf{-9.34}$ & $-9.35$ & $-12.00$ & $d_1$ \\
ARA ($C_1 \sim U(5,10)$) & $0.42$ & $0.58$ & $0.00$ & $-9.14$ & $\mathbf{-8.55}$ & $-12.00$ & $d_2$ \\
ARA-FPD $p^i_A \sim \text{Dir}(30,70,10)$ & $0.53$ & $0.47$ & $0.01$ & $-9.21$ & $\mathbf{-8.86}$ & $-11.92$ & $d_2$ \\
ARA-FPD $p^i_A \sim \text{Dir}(70,30,10)$ & $0.82$ & $0.17$ & $0.00$ & $\mathbf{-9.45}$ & $-9.81$ & $-11.94$ & $d_1$ \\
\end{tabular}
}
\end{table}

\noindent These results show the practical benefits of the ARA-FPD approach. Minimax forces the defender into worst-case resource allocation, often increasing defence budgets unnecessarily. Conventional ARA corrects this, but the defender has to precisely elicit the opponent's utility, which is complex and constitutes a critical point in the attack estimation, potentially making costly mistakes, as it may alter the optimal defence decision. 
ARA-FPD mitigates such elicitation burden, since placing uncertainty over an ideal probability is simpler than eliciting utilities in complex utility spaces. Additionally, the availability of the explicit FPD minimizer helps bypass the demanding nested simulation-optimization loops typical of standard ARA. This offers a viable way to tackle more complex, multi-stage problems that might otherwise lead to intractability under the conventional ARA framework, as we now show.


\section{ARA-FPD for multi-stage simultaneous defend-attack games} 
\label{sec:Rep_games}
\begin{figure}[pos=h]
  \centering
  \includegraphics[width=0.3\linewidth]{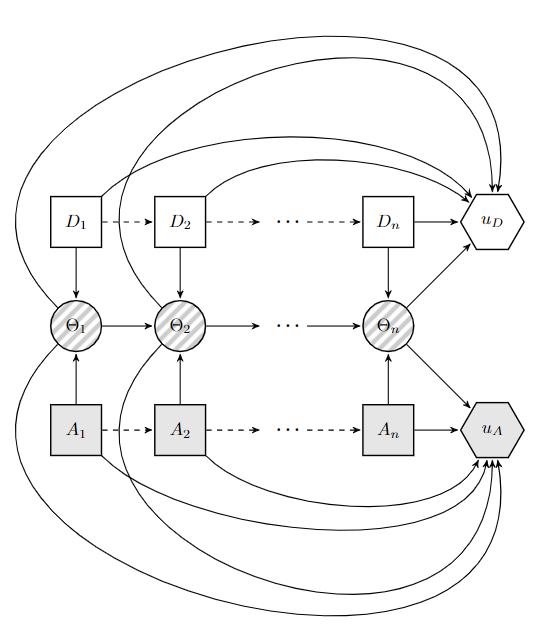} 
  \caption{{\small Template for basic \textit{n}-stage simultaneous defend-attack game BAID.}}
  \label{fig:n_sim_basic}
\end{figure}

Consider now repeated defend-attack interactions. Of the many possible variants,
let us focus on multi-stage simultaneous defend-attack games with incomplete information. Figure \ref{fig:n_sim_basic} specifies a BAID for an $n$-stage version of the interactions of interest. 
In the setup considered, at a first stage, the defender chooses $d_1 \in \mathcal{D}_1$, while the attacker simultaneously selects his attack $a_1 \in \mathcal{A}_1$. Following these actions, the uncertain outcome $\theta_1 \in \Theta_1$, influenced by both $a_1$ and $d_1$, is resolved and subsequently affects the uncertain $\theta_2 \in \Theta_2$, which also depends on actions $d_2 \in \mathcal{D}_2$ and $a_2 \in \mathcal{A}_2$. This pattern of simultaneous defend-attack decisions continues until the $n$-th stage. Finally, the utility of the defender is contingent upon all her decisions and uncertainties $\{\theta_j\}_{j=1}^n$, and, similarly, for the attacker.  

\begin{figure*}[pos = t]
    \centering
    \begin{subfigure}{0.4\textwidth}
        \centering
        \includegraphics[width=0.6\linewidth]{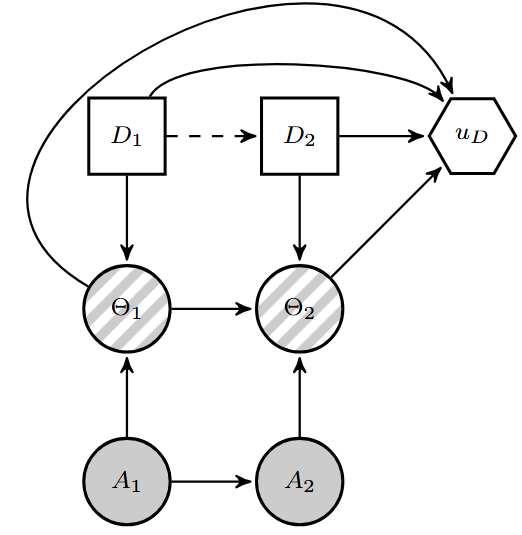}
        \caption{$D$'s influence diagram.}
        \label{AAA}
    \end{subfigure}
    \hfill
    \begin{subfigure}{0.4\textwidth}
        \centering
        \includegraphics[width=0.6\linewidth]{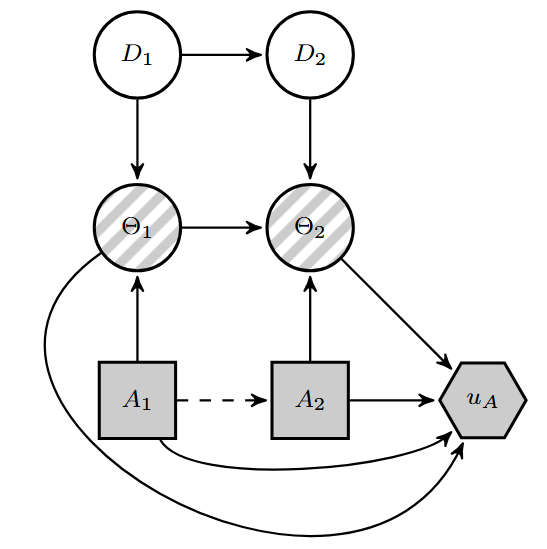}
        \caption{$A$'s influence diagram.}
        \label{BBB}
    \end{subfigure}
    
    \caption{\small{$D$'s and $A$'s influence diagrams for the 2-stage simultaneous defend-attack game.}}
    \label{fig:n_sim_DA_ids}
\end{figure*}

\noindent Appendix B shows how to solve a $2$-stage case with the aid of ARA by appealing to \citet{Shachter1986}'s ID reduction algorithm to the attacker's problem (to obtain the required attack forecasts) and to the defender's problem (to get the optimal defences). In doing so, as in the single stage case, we need to assess the attacker's random utility function, which is complex, and
 undertake complex computations. We mitigate both issues with the aid of ARA-FPD. 
For now, assume, as in Section~\ref{sec:State_action_controlSeq}, that the attacker aims to assess his sequence of attacks, designated $a(n) = (a_1,\dots,a_n) \in \mathcal{A}^n$, and outcomes, designated $\theta(n) = (\theta_1, \dots, \theta_n)\in \Theta^n$, that will result from the $n$-stage interaction.
Likewise, denote the sequence of defences as $d(n) = (d_1, \dots, d_n)$. 
FPD in the multi-stage case conceives the attacker as trying to make the joint density of all considered variables  $p_A(\theta(n),a(n),d(n))$ to approach his \textit{ideal} density $p^i_A(\theta(n),a(n),d(n))$ emulating the attacker preferences. Assume that such densities can be decomposed as
\begin{small}
\begin{equation*} \label{eq:joint_model}
p_A(\theta(n), a(n), d(n)) = \prod_{t=1}^n p_A(\theta_t,a_t,d_t \vert
\theta_{t-1}, a_{t-1}, d_{t-1})
\end{equation*}
\begin{equation}\label{eq:joint_ideal}
p^i_A(\theta(n), a(n), d(n)) = \prod_{t=1}^n p^i_A(\theta_t,a_t,d_t \vert
\theta_{t-1}, a_{t-1}, d_{t-1}). 
\end{equation}
\end{small}
with $p_A(\theta_1,a_1,d_1 \vert\theta_{0}, a_{0}, d_{0})=p_A(\theta_{1}, a_{1}, d_{1})$.
Moreover, for every $1 \leq t \leq n$ the joint distribution $p_A(\theta(n), a(n), d(n))$ can be factorized through \begin{small}
\begin{equation} \label{eq:factorisation_model}
 \begin{aligned}
  p_A(\theta_t,a_t,d_t \vert \theta_{t-1}, a_{t-1}, d_{t-1}) =
p_A(\theta_t \vert a_t,d_t, \theta_{t-1})
p_A(a_t) p_A(d_t)   .
 \end{aligned}   
\end{equation}
\end{small}
The factorization above relies on the following well-grounded assumptions \textbf{A1}-\textbf{A3},
 reflecting the BAID structure in Figure~\ref{fig:n_sim_basic}:
\begin{itemize}
    \item [\textbf{A1}] The outcome $\theta_t$ at time $t$, depends only on the outcome $\theta_{t-1}$ at time $t-1$, and the actions of both the attacker $a_t$ and the defender $d_t$.
    \item [\textbf{A2}] The attacker's action at time $t$ is not influenced by the previous outcome $\theta_{t-1}$ or the actions of either the defender or the attacker at time $t-1$.
    \item [\textbf{A3}] The defender's action at time $t$ is not influenced by the previous outcome $\theta_{t-1}$ or the actions of either the defender or the attacker at time $t-1$.
\end{itemize}
A factorisation similar to (\ref{eq:factorisation_model}) will be applied to the ideal distribution (\ref{eq:joint_ideal}) reflecting the attacker preferences
\begin{small}
\begin{equation} \label{eq:factorisation_preferences}
  p^i_A(\theta_t,a_t,d_t \vert \theta_{t-1}, a_{t-1}, d_{t-1}) =
p^i_A(\theta_t \vert a_t,d_t, \theta_{t-1})
p^i_A(a_t \vert d_t) p^i_A(d_t).   
\end{equation}
\end{small}
Under these assumptions, the attacker aims to obtain his optimal strategy, denoted $p^*_A(a_t)$. 
Following the FPD framework, for any time $t< n$, the optimal strategy will be determined using
\begin{small}
\begin{equation}\label{eq:FPD_multistage}
\begin{aligned}
 &\{p^*_A(a_t)\}_{t=1}^n = \argmin_{ \{p_A(a_t)\}_{t=1}^n\in {\Gamma}^n} KL [ p_A(\theta(n), a(n), d(n)) || p^i_A(\theta(n), a(n), d(n))], 
\end{aligned}
\end{equation}
\end{small}
where ${\Gamma^n}$ defines the space of admissible decision rules $p_A(a_t)$ available to the attacker. As before, such space might be constrained to a finite set of densities or a parametrized variational family.
Using factorizations (\ref{eq:factorisation_model}) and (\ref{eq:factorisation_preferences}), equation (\ref{eq:FPD_multistage}) can be rewritten
\begin{small}
\begin{equation} \label{eq:FPD_without DP}
    \begin{aligned}
    &\{p^*_A(a_t)\}_{t=1}^n =\\ & \argmin_{\{p_A(a_t)\}_{t=1}^n\in {\Gamma}^n} \;
         \int_{({\Theta, \cal{A}, {\cal{D}}})^n}  \prod_{t=1}^n
         p_A(\theta_t \vert a_t,d_t, \theta_{t-1})p_A(a_t) p_A(d_t)   \ln \left(\frac{  \prod_{t=1}^n \;p_A(\theta_t \vert a_t,d_t, \theta_{t-1})p_A(a_t) p_A(d_t)}{  \prod_{t=1}^n \;p^i_A(\theta_t \vert a_t,d_t, \theta_{t-1}) p^i_A(a_t \vert d_t) p^i_A(d_t)} \right) \dd\theta{(n)} \dd a{(n)} \dd d{(n)} = \\
        &\argmin_{\{p_A(a_t)\}_{t=1}^n\in {\Gamma}^n} \; \int_{({\Theta, \cal{A}, {\cal{D}}})^n}  \prod_{t=1}^n \; 
        p_A(\theta_{t} \vert a_{t},d_{t}, \theta_{t-1}) p_A(a_{t} ) p_A(d_{t}) \\
        & \times  \Bigg[ \ln \left(\frac{  \prod_{t=1}^n \;p_A(\theta_t \vert a_t,d_t, \theta_{t-1}) }{  \prod_{t=1}^n \;p^i_A(\theta_t \vert a_t,d_t, \theta_{t-1})} \right)+ \ln \left(\frac{  \prod_{t=1}^n \;p_A(a_t) }{  \prod_{t=1}^n \; p^i_A(a_t \vert d_t)}\right) \Bigg]
        \dd\theta{(n)} \dd a{(n)} \dd d{(n)} = \\ 
         & \argmin_{\{p_A(a_t)\}_{t=1}^n\in {\Gamma}^n} \; \int_{({\Theta, \cal{A}, {\cal{D}}})^n} \Bigg [  \prod_{t=1}^n \; 
        p_A(\theta_{t} \vert a_{t},d_{t}, \theta_{t-1})  \ln \left(\frac{  \prod_{t=1}^n \;p_A(\theta_t \vert a_t,d_t, \theta_{t-1}) }{  \prod_{t=1}^n \;p^i_A(\theta_t \vert a_t,d_t, \theta_{t-1})} \right) 
         \prod_{t=1}^n p_A(a_{t}) \prod_{t=1}^n p_A(d_{t})  \Bigg] \dd\theta{(n)} \dd a{(n)} \dd d{(n)} \\
        & +  \int_{{\cal{D}}^n} \prod_{\tau=1}^n \;p_A(d_{\tau})   KL\left(\prod_{t=1}^n \; p_A(a_{t}) \right\vert \left\vert \prod_{t=1}^n \; p^i_A(a_t \vert d_t)\right ) \dd d(n).
    \end{aligned}
\end{equation}
\end{small}

\noindent Equation~(\ref{eq:FPD_without DP}) implicitly uses the following assumption: as the attacker lacks specific information or preferences concerning the defender's actions, we assume that the attacker's ideal distribution for these actions matches their actual predictive distribution. Mathematically, this is expressed through $p^i_A(d_{t}) = p_A(d_{t})$. Using Assumptions~\textbf{A2}-\textbf{A3} the first term in (\ref{eq:FPD_without DP}) is rewritten as
\begin{small}
    \begin{equation}  \label{eq:auxil_FPD_without DP}
        \begin{aligned}
            &\int_{({\Theta, \cal{A}, {\cal{D}}})^n}  \prod_{t=1}^n 
        p_A(\theta_{t} \vert a_{t},d_{t}, \theta_{t-1}) p_A(a_{t}) p_A(d_{t}) \left[ \ln \left(\frac{  \prod_{t=1}^n p_A(\theta_t \vert a_t,d_t, \theta_{t-1}) }{  \prod_{t=1}^n p^i_A(\theta_t \vert a_t,d_t, \theta_{t-1})} \right) \right] \dd\theta{(n)} \dd a{(n)} \dd d{(n)} = \\
        & \sum_{t=1}^n  \int_{({{\Theta}^2, \cal{A}, {\cal{D}}})}  \ln \left(\frac{ p_A(\theta_t \vert a_t,d_t, \theta_{t-1}) }{ p^i_A(\theta_t \vert a_t,d_t, \theta_{t-1})} \right) \cdot p_A(\theta_{t}, a_{t}, d_{t}, \theta_{t-1}) \dd\theta_{t} \dd\theta_{t-1} \dd a_t \dd d_t = \\       
        & \int_{{{\cal{A}}^n}} \prod_{t=1}^n p_A(a_t)  \Biggl [ \sum_{t=1}^n \int_{({\Theta}^2,{\cal{D}})}  p_A(\theta_{t},\theta_{t-1}, d_{t}| a_{t}) \ln \left(\frac{ p_A(\theta_t \vert a_t,d_t, \theta_{t-1}) }{ p^i_A(\theta_t \vert a_t,d_t, \theta_{t-1})} \right)
         \; p_A(\theta_t) \; p_A(\theta_{t-1}) \; p_A(d_t)  \dd\theta_{t} \dd\theta_{t-1} \dd d_t \Biggr]  \dd a{(n)} = \\
        & \int_{{{\cal{A}}^n}} \prod_{t=1}^n p_A(a_t)\sum_{t=1}^n \; \mathbb{E}_{({\theta_{t-1}}, d_t)} \; KL\Bigl [p_A(\theta_t \vert a_t,d_t, \theta_{t-1})|| p^i_A(\theta_t \vert a_t,d_t, \theta_{t-1})\Bigr ]\dd a{(n)}.
    \end{aligned}
\end{equation}
\end{small}
Then, denoting the expected value in (\ref{eq:auxil_FPD_without DP}) 
\begin{small}
\begin{equation*}
 C(a_t) = \mathbb{E}_{({\theta_{t-1}}, d_t)} \; KL\Bigl [p_A(\theta_t \vert a_t,d_t, \theta_{t-1})|| p^i_A(\theta_t \vert a_t,d_t, \theta_{t-1})\Bigr ],
\end{equation*}
\end{small}
the optimal strategy (\ref{eq:FPD_multistage}) is
\begin{small}
\begin{equation} \label{eq:FPD_without DP_solution}
    \begin{aligned}
         &\{p^*_A(a_t)\}_{t=1}^n=  \argmin_{\{p_A(a_t)\}_{t=1}^n\in {\Gamma}^n} \;  \int_{{{\cal{A}}^n}} \prod_{t=1}^n p_A(a_t) \sum_{t=1}^n \; C(a_t)\dd a{(n)} +  \int_{{\cal{D}}^n} \prod_{\tau=1}^n \;p_A(d_{\tau})   KL\left(\prod_{t=1}^n \; p_A(a_{t}) \right\vert \left\vert \prod_{t=1}^n \; p^i_A(a_t \vert d_t)\right ) \dd d(n).
    \end{aligned}
\end{equation}
\end{small}
\noindent
Note that once the attacker's ideal does not depend on the defender's action, i.\ e.\ $p^i_A(a_t \vert d_t)=p^i_A(a_t)$, the optimal strategy (\ref{eq:FPD_without DP_solution}) adopts the simpler form 
\begin{small}
\begin{equation*}
    \begin{aligned}
         \{p^*_A(a_t)\}_{t=1}^n= & \argmin_{\{p_A(a_t)\}_{t=1}^n\in {\Gamma}^n} \;  \int_{{{\cal{A}}^n}} \prod_{t=1}^n p_A(a_t) \; \sum_{t=1}^n \; C(a_t)\dd a{(n)}  +  KL\left(\prod_{t=1}^n \; p_A(a_{t}) \right\vert \left\vert \prod_{t=1}^n \; p^i_A(a_t)\right ).
    \end{aligned}
\end{equation*}
\end{small}
Under this assumption, we obtain an explicit multi-stage solution version of (\ref{KL_DivExplicitsol}),

\begin{equation} \label{eq:n-stageFullSol}
    \prod_{t=1}^n p^*_A(a_t) \propto \prod_{t=1}^n p^i_A(a_t) \cdot \exp\{-\sum_{l=1}^nC(a_l)\},
\end{equation}
if \begin{small}    
$Z = \int_{\mathcal{A}^n} \prod_{t=1}^n p^i_A(a_t) \cdot \exp\{-\sum_{l=1}^nC(a_l)\}\dd a_1 \dots \dd{a_n}< \infty$.\end{small}
This formulation would allow us to simulate the full $n$-stage repeated game in just one step. However, we will not always be able to elicit the required ideal densities $p^i_A(\theta_t \vert a_t,d_t,\theta_{t-1})$ for $1 \leq t \leq n$, as it entails very involved reasoning.

A solution demanding less prior knowledge at the cost of increased computation follows.
Specifically, we only elicit the ideal density of the last stage outcome $p^i_A(\theta_n \vert a_n,d_n,\theta_{n-1})$, and obtain  the rest backwards, as with backward induction in game theory and decision analysis \citep{AUMANN19956}, solving the problem stage-wise. At the first stage in this simplified setting, $A$ seeks a sequence of attacks minimising (\ref{eq:FPD_multistage}), but, in addition to the previous assumptions, consider that \begin{small}  $p^i(\theta_{t-1} \vert a_{t-1},d_{t-1},\theta_{t-2}), p^i(a_{t-1})$\end{small} for $t<n$ are not available, {\em leaving them to its fate}. Under these conditions, (\ref{eq:FPD_without DP})  simplifies to  

\begin{small}
\begin{equation*}
    \begin{aligned}
       & \int_{({{\Theta}^2, \cal{A}, {\cal{D}}})}  \ln \left(\frac{ p_A(\theta_n \vert a_n,d_n, \theta_{n-1}) }{ p^i_A(\theta_n \vert a_n,d_n, \theta_{n-1})} \right)  p_A(\theta_{n}, a_{n}, d_{n}, \theta_{n-1}) \dd\theta_{n} \dd\theta_{n-1} \dd a_n \dd d_n + KL \left[p_A(a_n) \vert p^i_A(a_n) \right]=\\ &  \int_{\mathcal{A}} p_A(a_n) C(a_n) \dd a_n + KL \left[p_A(a_n) \vert p^i_A(a_n) \right],
    \end{aligned}
\end{equation*}
\end{small}
whose explicit solution is
\begin{equation}\label{eq:multiStageRecursive}
    p^*_A(a_n) \propto p_A^i(a_n) \cdot \exp\{C(a_n)\}.
\end{equation}
This explicit $n$-stage solution allows us to elicit the ideal density of the outcome at the previous state $\theta_{n-1}$. Specifically, the ideal density $p_{A}^i(\theta_{n-1})$ is assumed proportional to how close the (actual) density $p_A(\theta_n \vert a_n,d_n,\theta_{n-1})$ would be to his ideal density $p^i_A(\theta_n \vert a_n,d_n,\theta_{n-1})$ in the next stage: for every $\theta_{n-1} \in \Theta_{n-1}$, $p_{A}^i(\theta_{n-1})$ is
\begin{small}
\begin{equation} \label{eq:idealized_recursive}
    \begin{aligned}
    p_{A}^i&(\theta_{n-1}) \propto \exp\Bigg\{-\int_{\mathcal{A}}  p_A(a_n) \int_{\mathcal{D}} p_A(d_n) KL\left[p_A(\theta_n \vert a_n,d_n, \theta_{n-1}) \| p_A^i(\theta_n \vert a_n,d_n, \theta_{n-1}) \right] \dd a_n \dd d_n\Bigg\}.
    \end{aligned}
\end{equation}
\end{small}
Finally, we use this estimated ideal density of the outcome 
to obtain the attack density at stage $n-1$, and iterate this backward process recursively until we obtain all stages attack estimates.

Intuitively, (\ref{eq:n-stageFullSol}) yields the optimum attack probability over all stages, and such optimization is performed through all stages at once, jointly balancing them. Conversely, the proposed recursive formulation builds the full decision in a backward manner, taking into account only the expected value over the previous outcome, that is, the ideal density over $\theta$ at the $(t-1)$-th stage is selected to obtain better results at the $t$-th stage. 

As in Section \ref{sec:Seq_Games}, when modelling problems (\ref{eq:multiStageRecursive}) and (\ref{eq:idealized_recursive}) from the defender's perspective, we do not know the ideal densities $p^i_{A}(\theta_n\vert a_n, d_n)$, $p^i_A(a_t)$ nor the outcome densities $p_A(\theta_t \vert a_t,d_t, \theta_{t-1})$ for all $j \in \{1, \dots n\}$. Consequently, $D$ places uncertainty on them, using distributions $P_{A}(\theta_t \vert a_t,d_t, \theta_{t-1})$, $P^i_A(a_t)$, $P^i_{A}(\theta_n\vert a_n, d_n, \theta_{n-1})$ over densities, for all $t \in \{ 1, \dots, n\}$. Furthermore, since we are dealing with the simultaneous case, the defender has to also place uncertainty on the attacker’s beliefs over her action at each stage, namely $\Pi_{A}(d_t)$ for all $t \in \{ 1, \dots, n\}$. Therefore, the defender estimates the attacker’s density $p_A(a_t)$ for every stage, using $ F_t = (P_{A}(\theta_t \vert a_t,d_t, \theta_{t-1}),P^i_{A}(\theta_t \vert a_t,d_t, \theta_{t-1}), P ^i_A(a_t), \Pi_{A}(d_t)) .$ 
The random densities $P^i_{A}(\theta_t \vert a_t,d_t, \theta_{t-1})$ for $t\leq n-1$ are obtained recursively as follows.
Suppose we have estimated the attack density at stage $(t+1)$ and have a distribution $P^i_{A}(\theta_{t+1} \vert a_{t+1},d_{t+1}, \theta_{t})$ over the ideal densities.
Then, for every ideal density $\phi^k_{A}(\theta_{t+1} \vert a_{t+1},d_{t+1}, \theta_{t})$, define the corresponding random ideal density $\Phi^k_{A}(\theta_t \vert a_t,d_t, \theta_{t-1})$ at step $t$ as
\begin{small}
\begin{equation*} \label{eq:est_ideal}
\begin{aligned}
&\Phi^k_{A} (\theta_t) \propto \exp\Bigg\{-\int_{\mathcal{A}} p_{A}(a_{t+1})\cdot \int_{\mathcal{D}}  KL[P_{A}(\theta_{t+1} \vert d_{t+1},a_{t+1},\theta_t) \| \phi_{A}^k(\theta_{t+1} \vert d_{t+1},a_{t+1},\theta_t)] \dd d_{t+1} \dd a_{t+1}\Bigg\}.
\end{aligned}
\end{equation*}
\end{small}
Each ideal random density is approximated using MC simulation. We sample $\Big\{\pi_{A}^{k_1}(d_{t+1})\Big\}_{k_1=1}^{K_1} \sim \Pi_{A}(d_{t+1})$, and, for each of them, draw $K_2$ defences $\Big\{ d^{k_1}_{t+1,k_2}\Big\}_{k_2=1}^{K_2} \sim \pi_{A}^{k_1}(d_{t+1})$. Then, for every $ d^{k_1}_{t+1,k_2}$ we sample $$\Big\{ (p_{A}^l(\theta_{t+1} \vert d_{t+1},a_{t+1},\theta_t))\Big\}_{l=1}^L \sim (P_{A}(\theta_{t+1} \vert d_{t+1},a_{t+1},\theta_t))$$ and find, for each $l$,
\begin{small}
    \begin{equation} \label{eq:ideal_est}
    \begin{aligned}
        &\phi^{k,l,k_1,k_2}_{A} (\theta_t) = \exp\Bigg\{-\int_{\mathcal{A}} p_{A}(a_{t+1}) \cdot  KL[p^l_{A}(\theta_{t+1} \vert d^{k_1}_{t+1,k_2},a_{t+1},\theta_t) \| \phi_{A}^k(\theta_{t+1} \vert d^{k_1}_{t+1,k_2},a_{t+1},\theta_t) ] \dd a_{t+1}\Bigg\}.
        \end{aligned}
    \end{equation}
\end{small}
In the end, we estimate the $t^{th}$ stage ideal densities \begin{small}
$\{\hat{\phi}^k_{A}(\theta_t)\}_{k=1}^K$ \end{small}
\begin{small}
\begin{equation} \label{eq:ideal_est2}   
\hat{\phi}^k_{A}(\theta_t) = \frac{\sum_{k_1=1}^{K_1}\sum_{l= 1}^L\sum_{k_2=1}^{K_2}\phi^{k,l,k_1,k_2}_{A} (\theta_t)}{L\cdot K_1 \cdot K_2}.
\end{equation}
\end{small}
Starting from stage $t=n$, 
we estimate attack decisions at every stage through  $$(P_{A}(\theta_t \vert a_t,d_t, \theta_{t-1}),P^i_{A}(\theta_t \vert a_t,d_t, \theta_{t-1}), P^i_A(a_t), \Pi_{A}(d_t)) \sim F_t.$$ 
Similarly to problem (\ref{eq:MCMCSimultaneousStateactionControl}), we sample $\{\pi_{A}^{k_1}(d_t)\}_{k_1=1}^{K_1} \sim \Pi_{A}(d_t)$, and, from each, we draw $\{ d^{k_1}_{t,k_2}\}_{k_2=1}^{K_2} \sim \pi_{A}^{k_1}(d_t)$. Then, for each $ d^{k_1}_{t,k_2}$, we sample 
\begin{small}
\begin{equation*}
\begin{aligned}
    &\{ ((p^{i}_{A}(\theta_t \vert a_t,d^{k_1}_{t,k_2}, \theta_{t-1}),p^{i}_{A}(a_t \vert d^{k_1}_{t,k_2}),p_{A}(\theta_t \vert a_t,d^{k_1}_{t,k_2}, \theta_{t-1}))^l)\}_{l=1}^L \sim  (P^{i}_{A}(\theta_t \vert a_t,d^{k_1}_{t,k_2}, \theta_{t-1}),P^{i}_{A}(a_t),P_{A}(\theta_t \vert a_t,d^{k_1}_{t,k_2}, \theta_{t-1})),
    \end{aligned}
    \end{equation*}
\end{small}
and find, for each $l$,
\begin{small}
\begin{equation} \label{eq:att_est_multi_stage}
\begin{aligned}
    &p^{l,k_1,k_2}_{A}(a_t) \propto p^{il}_{A}(a_t)  \exp\Big\{-\mathbb{E}_{\Theta_{t-1}}\left[ KL \left[p_{A}^l(\theta_{t}\ \vert a_t,d^{k_1}_{t,k_2}, \theta_{t-1}) \| p^{il}_{A}(\theta_{t} \vert a_t,d^{k_1}_{t,k_2}, \theta_{t-1}) \right] \right]\Big\}.
\end{aligned}
\end{equation}
\end{small}
Importantly, when estimating $p^{l,k_1,k_2}_{A}(a_t)$, to calculate the expected value over $\Theta_{t-1}$ we consider $p(\theta_{t-1})$ to be uniformly distributed. This allows us to account for all previous outcomes equally when estimating the attacker's optimal solution at stage $t$.
Finally, we estimate the $t$-th stage attack density $\hat{p}_A(a_t)$, through the empirical frequencies

$$\hat{p}_A(a_t) = \frac{\sum_{k_1=1}^{K_1}\sum_{l= 1}^L\sum_{k_2=1}^{K_2}p^{l,k_1,k_2}_{A}(a_t)}{L\cdot K_1 \cdot K_2}.$$
The estimated $\hat{p}_A(a_t)$ is then used to approximate the ideal set of densities at stage  $(t-1)$ through (\ref{eq:ideal_est}) and (\ref{eq:ideal_est2}). This process continues recursively until stage 1, propagating uncertainty and obtaining all stages estimated attack densities, then used to calculate $D's$ optimal defences e.g. using \citet{Shachter1986} ID reduction algorithm. 

\subsection{Learning goals from experience}
Although there are contexts where experts are able to assess prior knowledge in the form of goals and preferences of the attacker, in certain contexts, it is conceivable that experts might not be able to provide the information about goals and preferences of the attacker, or even cases where the opponent's ideal goals and preferences vary over time.
Therefore, we now tackle how to update our knowledge about the opponent ideal based on experience. For this, we introduce a procedure to update knowledge after $m < n$ interactions in an $n$-stage simultaneous game. Intuitively, having observed the initial $m$ interactions, we evaluate which ideal densities are most likely to have generated the observed attacks and outcomes, accordingly updating our beliefs. 

Formally, suppose that we observe a trajectory    
$\{ (a_t,d_t, \theta_t)\}_{t=1}^m$, with $m<n$, and assume that $P^i_A(\theta_n,a_n \vert d_n,\theta_{n-1})$ is a distribution over a finite set of ideal densities,
    $\{ \phi_{n,A}^k (\theta_n, a_n\vert d_n, \theta_{n-1})\}_{k=1}^K$, which factorize as \begin{small}    
    $\phi_{n,A}^k(\theta_n\vert a_n,d_n, \theta_{n-1}) \phi_{n,A}^k(a_n)$\end{small}.
We seek the probability of the attacker employing a specific ideal $\phi_{n,A}^k$ given the observed trajectory. Applying Bayes' theorem and using the conditional independence assumptions \textbf{A1} and \textbf{A2}, this probability satisfies
\begin{equation*}
    \begin{aligned}
&p(\phi^k_{n,A} \vert a(m),d(m),\theta(m)) \propto p(\phi^k_{n,A} \vert d(m)) \cdot p_A(a(m),\theta(m) \vert \phi^k_{n,A},d(m)) =
p(\phi^k_{n,A}) \cdot \prod_{t=1}^m p_A(a_t \vert \phi^k_{n,A}) p_A(\theta_t \vert a_t,d_t,\theta_{t-1}).
\end{aligned}
\end{equation*} 
Denote by $\phi^k_{t,A}(\theta_t \vert a_t,d_t, \theta_{t-1})$ the recursively induced ideal densities at stage $t$. The density of an attack $a_t$, given an ideal density $\phi^k_{t,A}$, is
\begin{small}
\begin{equation*}
    \begin{aligned}
          &p(a_t\vert \phi^k_{n,A}) = \phi^k_{t,A}(a_t) \cdot \exp\Big\{- \int_{\mathcal{D}}KL \left[p_A(\theta \vert a_t,d, \theta_{t-1}) \| \phi^k_{t,A}(\theta \vert a_t, d,\theta_{t-1}) \right] \dd d \Big\}, 
    \end{aligned}
\end{equation*}
\end{small}
where we have removed the expectation over the outcomes at stage $(t-1)$, since we have already observed them. 
Under these conditions, for every ideal density, $\phi_{n,A}^k(a_n,\theta_n\vert d_n, \theta_{n-1})$, the defender estimates the probabilities $\{p(a_t\vert \phi^k_{n,A})\}_{t=1}^m$ using the distributions $(P_A(\theta_t \vert a_t,d_t,\theta_{t-1}),\Pi_A(d_t))$, to account for the uncertainty in the attacker's selection, as explained in Section \ref{sec:Seq_Games}. Specifically, the defender samples $K_1$ densities $\{\pi_{A}^{k_1}(d_t)\}_{k_1=1}^{K_1} \sim \Pi_{A}(d_t)$, and, from each, draws $K_2$ defences $\{ d^{k_1}_{t,k_2}\}_{k_2=1}^{K_2} \sim \pi_{A}^{k_1}(d_t)$. Then, for every $ d^{k_1}_{t,k_2}$, she samples $L$ times \begin{small}    
$\{ (p_{A}^l(\theta_t \vert a_t,d^{k_1}_{t,k_2},\theta_{t-1}))\}_{l=1}^L \sim (P_A(\theta_t \vert a_t,d^{k_1}_{t,k_2},\theta_{t-1}))$\end{small} and finds, for each $l$, 
\begin{small}
\begin{equation*}
    \begin{aligned}
    &p^{l,k_1,k_2}_{A}(a_t \vert \phi^k_{n,A}) \propto  \phi^k_{t,A}(a_t) \cdot \exp\left\{-KL \left[p_{A}^l(\theta \vert a_t,d^{k_1}_{t,k_2}, \theta_{t-1})\, \| \, \phi^k_{t,A}(\theta \vert a_t,d^{k_1}_{t,k_2}, \theta_{t-1}) \right]\right\}.
\end{aligned} 
\end{equation*}
\end{small}
The final estimated probabilities are obtained through the empirical frequencies

$$\hat{p}^{}_A(a_t \vert \phi^k_{n,A}) = \frac{\sum_{k=1}^{K_1}\sum_{l= 1}^L\sum_{k_2=1}^{K_2}\hat{p}^{l,k_1,k_2}_{A}(a_t \vert \phi^k_{n,A})}{L\cdot K_1 \cdot K_2}.$$
Once all probabilities are estimated, she updates her knowledge about the distribution $P^i_A(\theta_n,a_n \vert d_n, \theta_{n-1})$ through

\begin{equation} \label{eq:learn_exp}
p_{i+1}(\phi^k_{n,A}) = p_i(\phi^k_{n,A}) \cdot \prod_{t=1}^m \hat{p}_A(a_t \vert \phi^k_{n,A}) p_A(\theta_t \vert a_t,d_t,\theta_{t-1}),
\end{equation}
where $p_{i+1}(\phi_{n,A}^k)$ denotes the updated probability of $\phi_{n,A}^k$, after observing $i+1$ trajectories. This updated distribution is later used in the rest of the stages to sample from $P^i_{a,j}$, for all $j\in \{m+1, \dots, n \}$.

In summary, the framework provides a backward-forward algorithm. First, we construct the attacker's optimal decision (based on our beliefs) recursively backwards, deriving the ideal densities $\phi^k_{t,A}$ for every stage, starting from the terminal stage $n$. Then, the learning mechanism operates in a forward manner, taking the observed trajectory $\{ (a_t,d_t,\theta_t)\}_{t=1}^m$ and iteratively updating the defender's posterior beliefs about the attacker's actual goals. 
\section{Examples}\label{sec:Exp}
This section shows the results of applying the ARA-FPD methodology from Section \ref{sec:Rep_games} to an air combat problem, modeled as a two-stage version of the BAID shown in Figure \ref{fig:n_sim_basic}.
Consider a scenario in which a defender manages an aerial defence system to protect a strategic infrastructure targeted by an attacker.
We support the defender in managing infrastructure protection through two decisions. At the first defence stage ($D_1$), $D$ chooses the level of anti-aircraft missile deployment (minimal $d_1^1$; moderate $d_1^2$; or substantial $d_1^3$) while the attacker simultaneously decides within the attack set ($A_1$) whether to conduct no attack ($a_1^1$), a low-intensity drone strike ($a_1^2$), or a high-intensity drone strike ($a_1^3$), aimed at weakening those defences. The outcome of this interaction is $\Theta_1$, indicating whether the drones succeed in neutralizing the anti-aircraft system. In the second defence stage ($D_2$), $D$ determines how many defensive drones to deploy to protect the infrastructure (no support $d_2^1$; minimal $d_2^2$; or substantial $d_2^3$) while the attacker decides in the attack set ($A_2$) the number of drones to be sent targeting the infrastructure (no drones ($a_2^1$), low-amount ($a_2^2$), or high-amount ($a_2^3$)). The combination of these decisions and the previous outcome $\Theta_1$ determines whether the infrastructure is ultimately destroyed, reflected through by the random variable $\Theta_2$.

\paragraph{Models.} The Defender’s beliefs about $\Theta_1$, conditional on $d_1$ and $a_1$, are modelled with the probabilities in Table \ref{tab:succes_miss_prob}a. Additionally, the probabilities that her infrastructure is destroyed ($\Theta_2$), conditional on $a_2$, $d_2$ and whether the attack on the anti-air missiles was successful are outlined in Tables \ref{tab:succes_miss_prob}b and \ref{tab:succes_miss_prob}c.

\begin{table}[pos = h]
    \centering
    \caption{Probabilities of successful attack in first stage (a) and second stage (b) and (c).}
    \begin{subtable}{0.3\columnwidth}
        \centering
        \caption{$p_D(\Theta_1$=$1|d_1,a_1)$}
        \begin{tabular}{c|ccc}
                              & $a^1_1$ & $a_1^2$ & $a_1^3$ \\ \hline
\multicolumn{1}{c|}{$d_1^1$} & 0.00    & 0.70    & 0.80    \\
\multicolumn{1}{c|}{$d_1^2$} & 0.00    & 0.35    & 0.70    \\
\multicolumn{1}{c|}{$d_1^3$} & 0.00    & 0.10    & 0.20    \\ 
\end{tabular}
        \label{tab:sub_first}
    \end{subtable}
    \hspace{0.5cm}
    \begin{subtable}{0.3\columnwidth}
        \centering
         \caption{$p_D(\Theta_2$=$1|d_1,a_1,\Theta_1 $=$1)$}
        \begin{tabular}{c|ccc}
                              & $a^1_2$ & $a_2^2$ & $a_2^3$ \\ \hline
\multicolumn{1}{c|}{$d_2^1$} & 0.00    & 0.85    & 0.95    \\
\multicolumn{1}{c|}{$d_2^2$} & 0.00    & 0.40    & 0.70    \\
\multicolumn{1}{c|}{$d_2^3$} & 0.00    & 0.20    & 0.35    \\ 
\end{tabular}
        \label{tab:sub_second}
    \end{subtable}
    \hfill
    \begin{subtable}{0.3\columnwidth}
        \centering
        \caption{$p_D(\Theta_2$=$1|d_1,a_1,\Theta_1 $=$0)$}
       \begin{tabular}{c|ccc}
                              & $a^1_2$ & $a_2^2$ & $a_2^3$ \\ \hline
\multicolumn{1}{c|}{$d_2^1$} & 0.00    & 0.25    & 0.35    \\
\multicolumn{1}{c|}{$d_2^2$} & 0.00    & 0.15    & 0.25    \\
\multicolumn{1}{c|}{$d_2^3$} & 0.00    & 0.05    & 0.10    \\ 
\end{tabular}
        \label{tab:sub_third}
    \end{subtable}
    \label{tab:succes_miss_prob}
\end{table}

\paragraph{Defender Modelling of  Attacker's Beliefs.} The defender models the attacker's beliefs about $\Theta_1 = 1$ as $P_A(\Theta_1|d_1,a_1) \sim \beta e (\sigma_1,\sigma_2)$, with parameters $(\sigma_1, \sigma_2)$ for each $d_1$ and $a_1$ as in Table \ref{tab:beta_succes_miss_prob}a. Similarly, the defender assesses $A$'s beliefs about $\Theta_2 = 1$ as $P_A(\Theta_2|d_2,a_2, \theta_1) \sim \beta e(\sigma_3,\sigma_4)$, with parameters $(\sigma_3, \sigma_4)$ in Tables \ref{tab:beta_succes_miss_prob}b and \ref{tab:beta_succes_miss_prob}c. A single number (0 or 1) for the corresponding combination of arguments in a table means that a degenerate distribution is used there. Additionally, $D$ models $A$'s assumptions regarding her decisions to install anti-air missiles through a Dirichlet distribution $P_A(d_1) \sim \text{Dir}(1, 75, 25)$. As for the attacker's beliefs about $D_2$, the defender uses $P_A(d_2 \vert \Theta_1=1) \sim \text{Dir}(2,18,80)$ and $P_A(d_2 \vert \Theta_1=0) \sim \text{Dir}(2,33,65)$. 

$D$'s beliefs about the attacker's ideal densities
$p^i_2(\theta_2|a^i_2,d^j_2,\theta_1)$ are $p^i_2(\theta_2=1|a^i_2,d^j_2,\theta_1) = 1,  \forall i,j \in \{1,2,3\},\theta_1 \in \{0,1\}$. The ideal attack probabilities (preferences) are considered to be in the set  \begin{small}    
$\Big\{(p_1^{i,k},p_2^{i,k}) \Big\}_{k=1}^3$\end{small} where $(p_1^i,p_2^i)^k$ are in Table \ref{tab:believed_id_att_probs}.
\begin{table}[pos = h]
    \centering
    \caption{Beta parameters for $D$'s modelling of $A$'s beliefs on $\Theta_1 = 1$, $\Theta_2 = 1$.}
    \begin{subtable}{0.3\columnwidth}
        \centering
        \caption{$P_A(\Theta_1$=$1|d_1,a_1)$}
       \begin{tabular}{c|ccc}
                              & $a^1_1$ & $a_1^2$ & $a_1^3$ \\ \hline
\multicolumn{1}{c|}{$d_1^1$} & 0 & (21,9) & (30,5) \\
\multicolumn{1}{c|}{$d_1^2$} & 0 & (16,24) & (18,4) \\
\multicolumn{1}{c|}{$d_1^3$} & 0 & (4,18) & (10,10) \\ 
\end{tabular}
        \label{tab:sub_first}
    \end{subtable}
    \hfill 
    \begin{subtable}{0.3\columnwidth}
        \centering
         \caption{$P_A(\Theta_2$=$1|d_2,a_2,\Theta_1 $=$1)$}
        \begin{tabular}{c|ccc}
                              & $a^2_1$ & $a_2^2$ & $a_2^3$ \\ \hline
\multicolumn{1}{c|}{$d_2^1$} & 0 &1 & 1 \\
\multicolumn{1}{c|}{$d_2^2$} & 0 & (5,5) & (24,4) \\
\multicolumn{1}{c|}{$d_2^3$} & 0 & (12,48) & (48,12) \\ 
\end{tabular}
        \label{tab:sub_second}
    \end{subtable}
    \hfill
        \begin{subtable}{0.3\columnwidth}
        \centering
        \caption{$P_A(\Theta_2$=$1|d_1,a_1,\Theta_1 $=$0)$}
       \begin{tabular}{c|ccc}
                              & $a^1_2$ & $a_2^2$ & $a_2^3$ \\ \hline
\multicolumn{1}{c|}{$d_2^1$} & 0 & (12,18) & (18,12) \\
\multicolumn{1}{c|}{$d_2^2$} & 0 & (6,24) & (9,21) \\
\multicolumn{1}{c|}{$d_2^3$} & 0 & (3,27) & (2,24) \\ 
\end{tabular}
        \label{tab:sub_third}
    \end{subtable}
    \label{tab:beta_succes_miss_prob}
\end{table}
\begin{table}[pos = h]
    \centering
    \caption{Defender's believed ideal attack probabilities.}
    \begin{subtable}{0.3\columnwidth}
        \centering
        \caption{Stage 1}
        \begin{tabular}{c|ccc}
            & $a_1^1$ & $a_1^2$ & $a_1^3$ \\ \hline
            \multicolumn{1}{c|}{$p_1^{i,0}(a_1)$} & 0.00 & 0.5 & 0.5 \\
            \multicolumn{1}{c|}{$p_1^{i,1}(a_1)$} & 0.0 & 0.6 & 0.4 \\
            \multicolumn{1}{c|}{$p_1^{i,2}(a_1)$} & 0.00 & 0.3 & 0.7 \\ 
        \end{tabular}
        \label{tab:sub_first_stage}
    \end{subtable}%
    \hspace{0.5cm}
    \begin{subtable}{0.3\columnwidth}
        \centering
        \caption{Stage 2}
        \begin{tabular}{c|ccc}
            & $a_2^1$ & $a_2^2$ & $a_2^3$ \\ \hline
            \multicolumn{1}{c|}{$p_2^{i,0}(a_2)$} & 0.1 & 0.5 & 0.4 \\
            \multicolumn{1}{c|}{$p_2^{i,1}(a_2)$} & 0.1 & 0.4 & 0.5 \\
            \multicolumn{1}{c|}{$p_2^{i,2}(a_2)$} & 0.0 & 0.2 & 0.8 \\ 
        \end{tabular}
        \label{tab:sub_second_stage}
    \end{subtable}
    \label{tab:believed_id_att_probs}
\end{table}

\paragraph{Defender's utility.} The defender preferences are modeled with a constant risk-averse utility function \begin{small}   $$u_D(d_1,a_1,\theta_1,d_2,a_2,\theta_2) = - \textrm{exp} \Big(\rho_D \cdot c_D(d_1,a_1,\theta_1,d_2,a_2,\theta_2)\Big),$$ \end{small} where $\rho_D$ is her risk aversion coefficient and $c_D$ denotes the monetary costs related to $d_1, a_1, \theta_1,d_2,a_2$ and $\theta_2$. Specifically, assume $c_D(a_1,d_1,a_2,d_2,\theta_1,\theta_2)= f_D(\theta_2) +f_D(\theta_1) + (1+f_D(a_1))f_D(d_1)+(1+f_D(a_2))f_D(d_2)$, where $f_D$ is a monetary value function displayed in Table \ref{fig:f_d_f_a}. 

\begin{table*}[pos = h]
\centering
\caption{Values of $f_D$ for decisions and uncertainties.}
\resizebox{0.8\textwidth}{!}{
\begin{tabular}{c|cccccc|cc|cccccc|cc}
                            & $d_1^1$ & $d_1^2$ & $d_1^3$ & $a_1^1$ & $a_1^2$ & $a_1^3$ & $\Theta_1$=0 & $\Theta_1$=1 & $d_2^1$ & $d_2^2$ & $d_2^3$ & $a_2^1$ & $a_2^2$ & $a_2^3$ & $\Theta_2$=0 & $\Theta_2$=1 \\ \hline
\multicolumn{1}{c|}{$f_D$} & 0.0     & 0.3     & 0.6     & 0.0     & 0.4     & 0.5     & 0.0          & 2.0          & 0.0     & 0.3     & 1     & 0.0     & 0.5     & 0.8    & 0.0          & 7.5          \\

\end{tabular}
}
\label{fig:f_d_f_a}
\end{table*}

\subsection{2-stage interaction} \label{sec:2sgame}
First, we use the scenario to showcase how the ARA-FPD approach operates in a $2$-stage setting, presenting the estimated attacker's probabilities and optimal defences obtained with the approach in Section \ref{sec:Rep_games}. The optimal decision for the first stage is to build a substantial anti-air defence ($d_1^* = d_1^3)$. As for the second stage, the best response is to protect the infrastructure with a moderate defence ($d_2^* = d_2^2$). Regarding the probabilistic forecasts about the attacker, our approach suggests a moderate attacker in the first stage, being $p(a_1) = (0.00, 0.45, 0.55)$, whereas the second stage forecasts suggest a slightly more conservative attacker, $p(a_2) = (0.11, 0.43, 0.46)$. 

\subsection{Multi-stage interactions with experience learning}
Let us extend now the example to a multi-stage setting where the two-stage interaction in Section \ref{sec:2sgame} is repeated $k$ times ($2k$ stages), while the defender learns from every two interactions with the attacker, i.e. every time the two-stage game finishes. Each of these repetitions constitutes an episode. This repeated procedure models a scenario where the defender protects multiple similar infrastructures. Therefore, an outcome in a specific episode serves to update the defender's beliefs, allowing her to adapt the defensive strategies for subsequent attacks on remaining locations.
We use two kinds of attackers to compare the effectiveness of the learning procedure.
\vspace{0.1in}

\noindent \textbf{Stationary attacker.}
Let us begin by analysing how the defender adapts its knowledge when facing an attacker modelled through the ARA-FPD framework under a fixed ideal probability. The attacker is assumed to select the density closest to its ideal, while keeping that ideal fixed throughout the episodes. The ideal outcomes are
$$p^i_1(\theta_1=1|a^i_1,d^j_1) = 1,  \forall i,j \in \{1,2,3\}$$
$$p^i_2(\theta_2=1|a^i_2,d^j_2,\theta_1) = 1,  \forall i,j \in \{1,2,3\},\theta_1 \in \{0,1\},$$
while the ideal attack probabilities are
\begin{equation} \label{eq:ideal_att_probs}
\begin{aligned}
    \hspace{4cm}p_1^i(a_1) &= (0,0.4,0.6), \hspace{0.5cm} p_2^i(a_2) = (0.1,0.5,0.4).
\end{aligned}
\end{equation}
Once the optimal attacks are determined, the resulting outcomes are computed using the probabilities in Table~\ref{tab:succes_miss_prob}.
When the defences, attacks and outcomes are observed, the defender updates her beliefs on the attacker's ideal prior distribution through (\ref{eq:learn_exp}). Note that since we have already observed the actions adopted, the defender's modelling of the attacker's belief about her second decision is more accurate, as in Table \ref{tab:dir_par_aircombat}. The defender starts with a uniform prior distribution over the set of ideals. We run the simulation for 200 episodes, with 100 distribution and defence samples per episode and 1000 $P_A$ samples, i.e. $K_1=K_2=100$ and $L=1000$ in (\ref{eq:att_est_multi_stage}). We use $\alpha= 0.2$ to smooth the probabilities.
\begin{table}[pos = h]
    \centering
    \caption{Dirichlet parameters for $D$'s modelling of $A$'s beliefs on $D_2$.}
    \begin{subtable}{0.44\columnwidth}
        \centering
        \caption{$P_A(d_2|d_1,a_1,\Theta_1=1)$}
\begin{tabular}{c|ccc}
                           & $a_1^1$    & $a_1^2$      & $a_1^3$      \\ \hline
\multicolumn{1}{c|}{$d_1^1$}&(15, 35, 50) & (5,25,70) & (3,7,90) \\
\multicolumn{1}{c|}{$d_1^2$}&(15, 14, 60) & (5,20,80) & (2,6,92) \\
\multicolumn{1}{c|}{$d_1^3$}& (10, 20, 70)& (5,15,85) & (2,20,90)  \\ 
\end{tabular}
        \label{tab:first_dataset}
    \end{subtable}%
    \begin{subtable}{0.44\columnwidth}
        \centering
        \caption{$P_A(d_2|d_1,a_1,\Theta_1=0)$}
\begin{tabular}{c|ccc}
                              & $a_1^1$         & $a_1^2$          & $a_1^3$           \\ \hline
\multicolumn{1}{c|}{$d^1_1$} & (20,40,60) & (10,40,70)    & (5,20,80)     \\
\multicolumn{1}{c|}{$d^2_1$} & (10,40,60)  & (10,30,60) & (5,30,80)  \\
\multicolumn{1}{c|}{$d^3_1$} & (20,30,40)  & (10,30,80)  & (5,35,90) \\ 
\end{tabular}
        \label{tab:second_dataset}
    \end{subtable}
\label{tab:dir_par_aircombat}
\end{table} 

\begin{figure}[pos = h]
    \centering
    \begin{subfigure}[c]{0.43\textwidth}
        \centering
        \includegraphics[width=\linewidth]{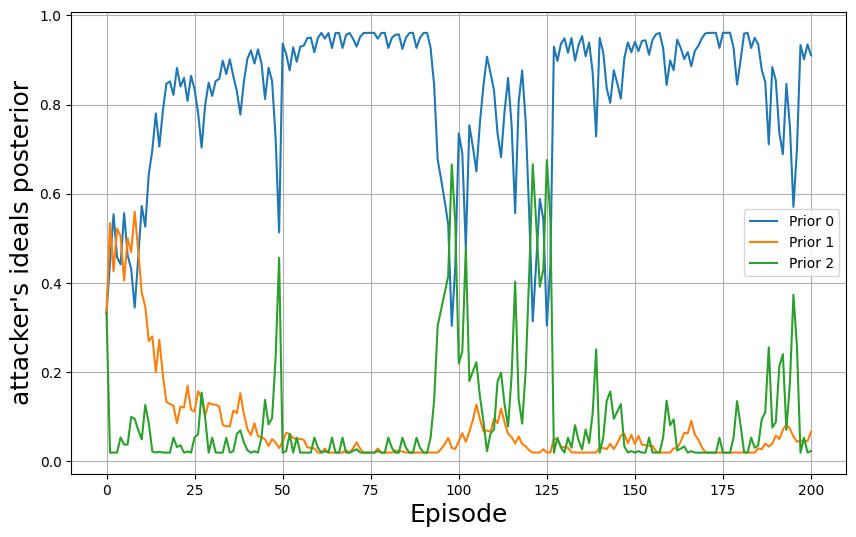}
        \caption{ Evolution of defender's posterior distribution over attacker's ideals (stationary).}
        \label{fig:id_prior_fixed_main}
    \end{subfigure}
    \hfill
    \begin{minipage}[c]{0.53\textwidth}
        \centering
        \begin{subfigure}[b]{\linewidth}
            \centering
            \includegraphics[width=0.9\linewidth]{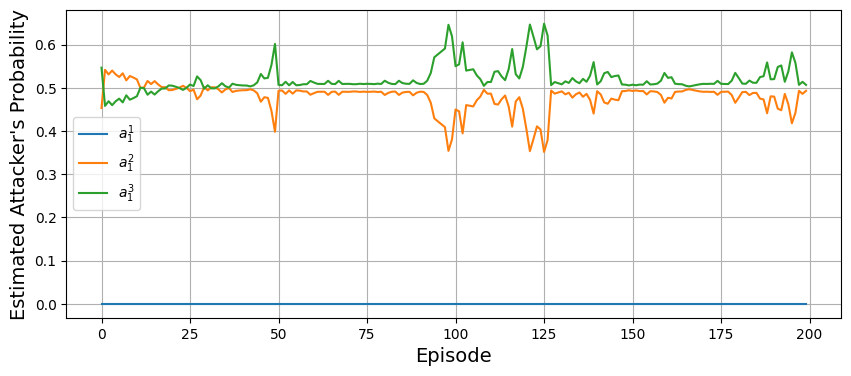}
            \caption{Estimated attacker probabilities for a fixed attacker. Stage 0.}
            \label{fig:probfixedS0}
        \end{subfigure}
        \begin{subfigure}[b]{\linewidth}
            \centering
            \includegraphics[width=0.9\linewidth]{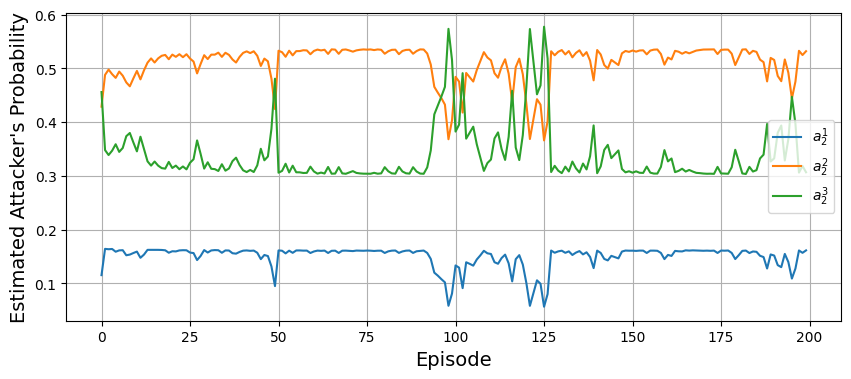}
            \caption{Estimated attacker probabilities for a fixed attacker. Stage 1.}
            \label{fig:probfixedS1}
        \end{subfigure}
    \end{minipage}
    \caption{Defender beliefs about attacker ideals (left) and attack probabilities (right) for the stationary attacker throughout the episodes.}
    \label{fig:fixed_attacker_main}
\end{figure}

\begin{figure}[pos = htpb]
    \centering
    \begin{subfigure}[b]{0.32\textwidth}
        \centering
        \includegraphics[width=\textwidth]{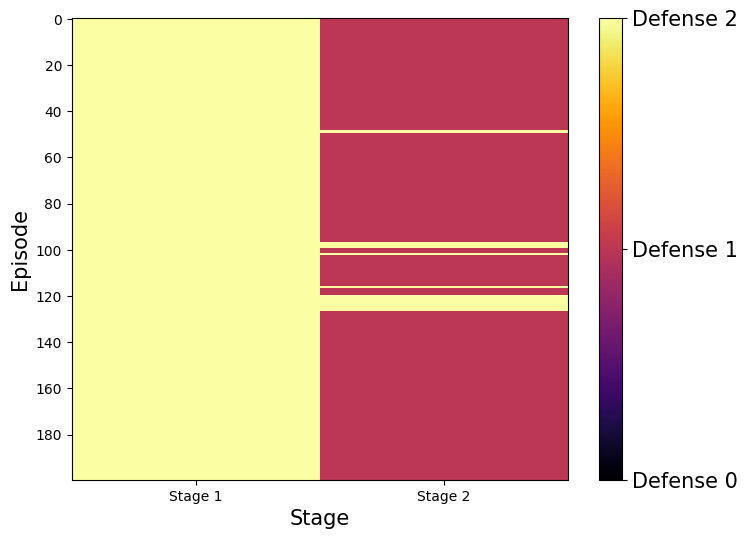}
        \caption{Defences frequency.}
        \label{fig:defense_heatmap_fixed}
    \end{subfigure}
    \begin{subfigure}[b]{0.32\textwidth}
        \centering
        \includegraphics[width=\textwidth]{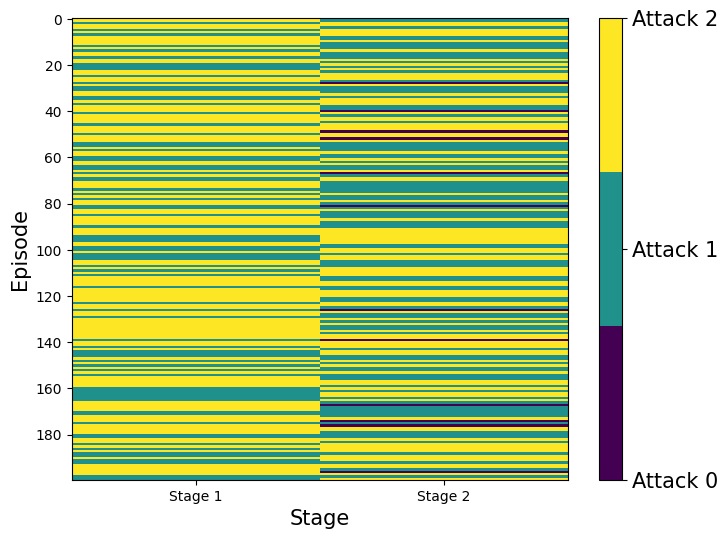}
        \caption{Attack frequency.}
        \label{fig:attack_freq_fixed}
    \end{subfigure}
    \begin{subfigure}[b]{0.32\textwidth}
        \centering
        \includegraphics[width=\textwidth]{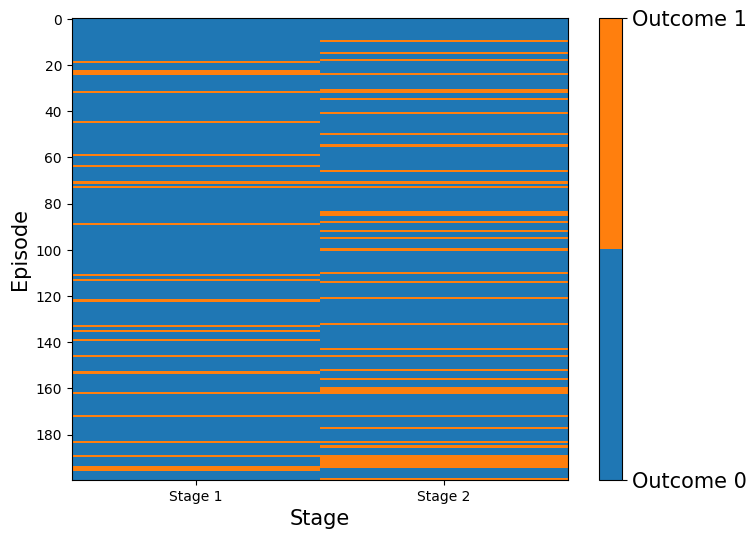}
        \caption{Defender outcomes.}
        \label{fig:outcome_freq_fixed}
    \end{subfigure}
    \caption{Overview of defences, attacks, and outcomes per episode and stage for a fixed attacker.}
\label{fig:combined_defense_attack_outcome_fixed}
\end{figure}

\noindent Observe that, throughout the episodes, the optimal defences generally remain as $d_1^3$ and $d_2^2$. By the final episode, the estimated attack probabilities in the first stage closely resemble the 2-stage game results, yielding $p(a_1) = (0.00, 0.49, 0.51)$. Conversely, in the second stage, the defender infers a more conservative attacker, adjusting her estimate to $p(a_2) = (0.16, 0.53, 0.31)$. Figure \ref{fig:id_prior_fixed_main} illustrates the evolution of the defender's posterior distribution over the set of ideals, while Figures \ref{fig:probfixedS0} and \ref{fig:probfixedS1} show the estimated attack probabilities across all episodes. Finally, Figure \ref{fig:combined_defense_attack_outcome_fixed} summarizes the observed actions (defences and attacks) and outcomes. The figure is divided into Stage 1 (left) and Stage 2 (right), with each plot detailing the chosen defences, executed attacks, and resulting outcomes over time.
\vspace{0.1in}


 \noindent \textbf{Adaptive attacker.}
Consider now an opponent modelled as following an ARA-FPD approach but changing his ideal probability throughout the episode based on previous interactions. We use the same settings as above, with the ideal probability in  (\ref{eq:ideal_att_probs}). The difference now is that, after each episode, the attacker revisits the ideal probability at every stage: if $\theta_i=0,  i \in \{1,2\}$, then the attacker increases the ideal probability of the aggressive attack, for every defence, in stage $i$; if $\theta_i=1,  i \in \{1,2\}$, then the attacker increases the ideal probability of the moderate attack with probability $0.5$ for stage $i$. 
The defender starts with a uniform distribution over the set of ideals. We run the simulation for 200 episodes, with 100 distribution and defence samples per episode, and 1000 $P_A$ samples (i.e. $K_1=K_2=100$, $L=1000$ in (\ref{eq:att_est_multi_stage})). Again, we use a smoothing parameter $\alpha= 0.2$. 

\begin{figure}[pos = h]
    \centering
    \begin{subfigure}[c]{0.43\textwidth}
        \centering
        \includegraphics[width=\linewidth]{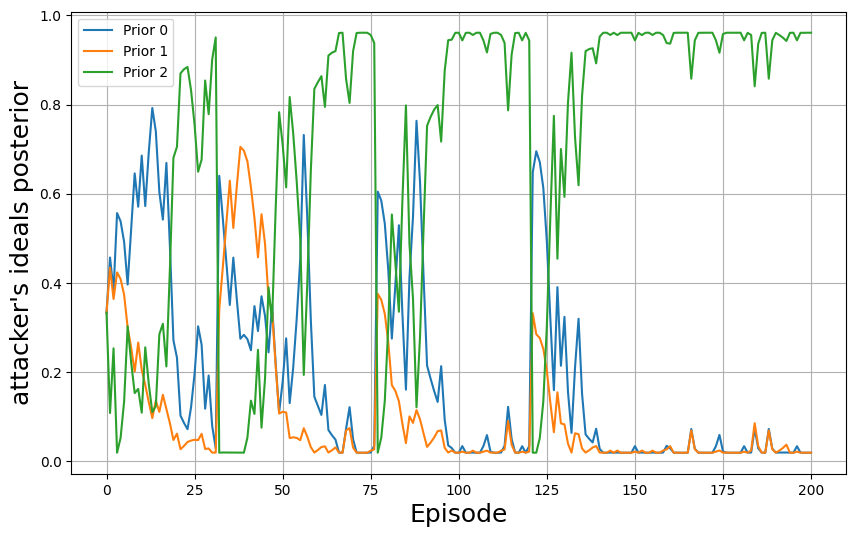}
        \caption{Evolution of defender's posterior distribution over attacker's ideals (adaptive).}
        \label{fig:id_prior_adaptive_main}
    \end{subfigure}
    \hfill
    \begin{minipage}[c]{0.53\textwidth}
        \centering
        \begin{subfigure}[b]{\linewidth}
            \centering
            \includegraphics[width=0.9\linewidth]{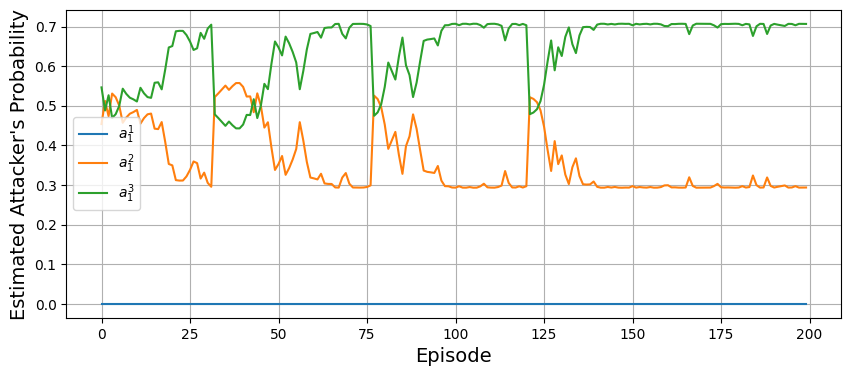}
            \caption{Estimated attacker probabilities for an adaptive attacker. Stage 0.}
            \label{fig:probadapS0}
        \end{subfigure}
        
        \vspace{0.3cm} 

        \begin{subfigure}[b]{\linewidth}
            \centering
            \includegraphics[width=0.9\linewidth]{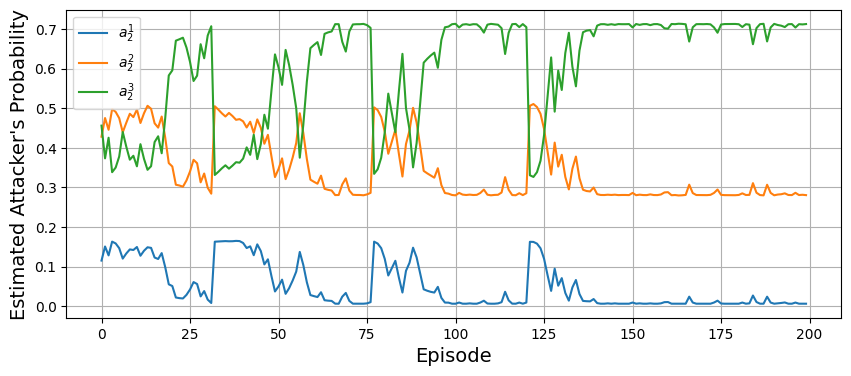}
            \caption{Estimated attacker probabilities for an adaptive attacker. Stage 1.}
            \label{fig:probAdapS1}
        \end{subfigure}
    \end{minipage}

    \vspace{10pt} 
    \caption{Defender beliefs about attacker ideals (left) and attack probabilities (right) for the adaptive attacker throughout the episodes.}
    \label{fig:adaptive_attacker_main}
\end{figure}

\begin{figure*}[pos = htpb]
    \centering
    \begin{subfigure}[b]{0.32\textwidth}
        \centering
        \includegraphics[width=\textwidth]{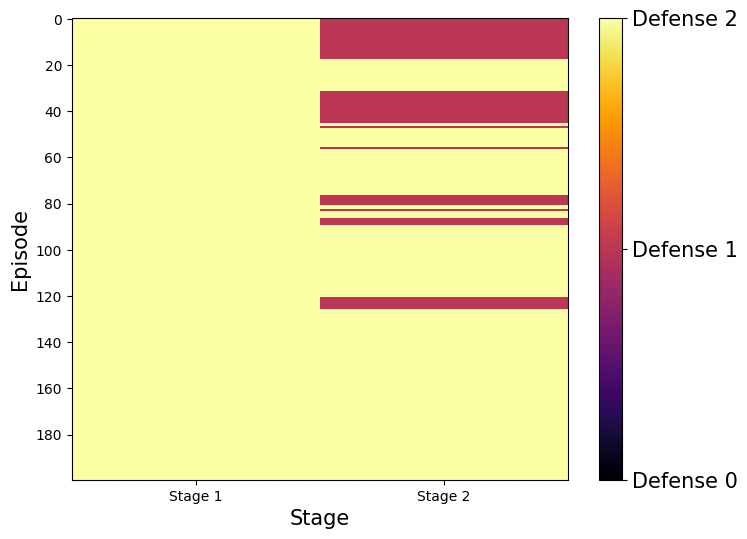}
        \caption{Defences frequency.}
        \label{fig:defense_heatmap_adap}
    \end{subfigure}
    \begin{subfigure}[b]{0.32\textwidth}
        \centering
        \includegraphics[width=\textwidth]{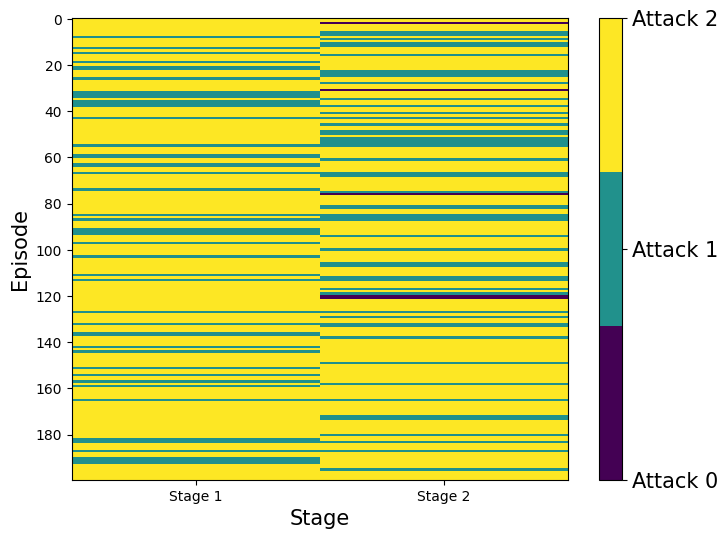}
        \caption{Attack frequency.}
        \label{fig:attack_freq}
    \end{subfigure}
    \begin{subfigure}[b]{0.32\textwidth}
        \centering
        \includegraphics[width=\textwidth]{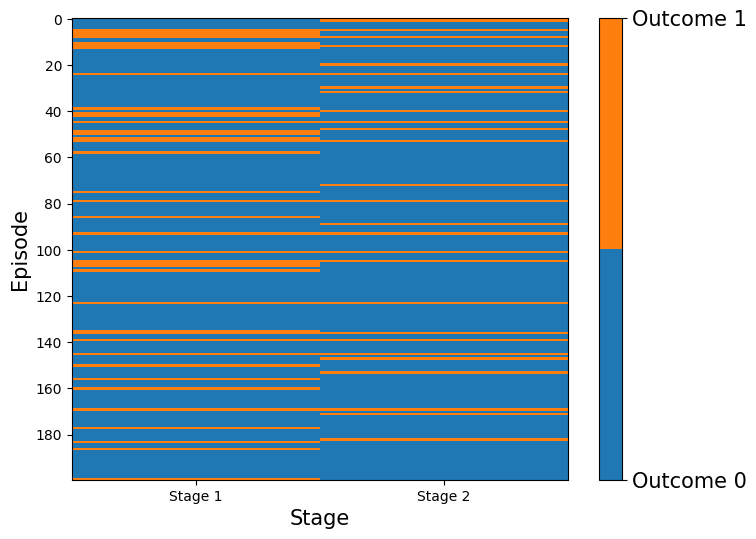}
        \caption{Defender outcomes.}
        \label{fig:outcome_freq_adaptive}
    \end{subfigure}
    \caption{Overview of defences, attacks, and outcomes per episode and stage for an adaptive attacker.}
\label{fig:combined_defense_attack_outcome}
\end{figure*}

The results indicate that while the defender initially selects the optimal defences $d_1^3$ and $d_2^2$, she rapidly shifts to a more conservative approach, with $d_1^3$ and $d_2^3$ the optimal defences in the final episodes. Correspondingly, for the last episode, the estimated attack probabilities reflect a significantly more aggressive opponent, reaching \begin{small}$p(a_1) = (0.00, 0.29, 0.71)$\end{small} in the first stage and \begin{small}$p(a_2) = (0.02, 0.28, 0.70)$\end{small} in the second.
Figure \ref{fig:id_prior_adaptive_main} illustrates the evolution of the posterior distribution over the attacker's ideals. While the conservative priors initially dominate the defender's beliefs, observe how the posterior mass shifts towards the aggressive prior as the episodes progress. Figures \ref{fig:probadapS0} and \ref{fig:probAdapS1} track the estimated attack probabilities over time, capturing how the defender adapts to the increasing aggressive behaviour of the attacker. Finally, Figure \ref{fig:combined_defense_attack_outcome} summarizes the observed actions and outcomes in the scenario. Notably, it shows how the attacker's aggressive behaviour translates into a greater success in the first stage, but also portrays how the defender successfully adapts by deploying more substantial defences to neutralize the threat in the second stage.

\subsection{Discussion}
We end up by comparing the dynamics across different experiments. In the repeated two-stage games, the defender's posterior over the attacker's ideals behaves differently depending on the opponent's behaviour. When facing a stationary attacker with fixed ideal probabilities, the defender's belief initially oscillates between both conservative priors (priors 0 and 1), while the aggressive prior's (prior 2) weight quickly decays towards zero. Ultimately, prior 0 captures the majority of the posterior mass. Interestingly, when the defender detects a sequence of strong attacks (Figure \ref{fig:attack_freq_fixed}), the weight of the aggressive prior temporarily increases (Figure \ref{fig:id_prior_fixed_main}). This induced uncertainty directly affects the estimated attack probabilities in the second stage, which also start to oscillate (Figure \ref{fig:probfixedS1}), resulting in a temporary shift towards a substantial defence ($d_2^3$) during that period (Figure \ref{fig:defense_heatmap_fixed}).

In contrast, when facing the adaptive attacker, the conservative priors dominate only during the early episodes. As the episodes progress, the posterior distribution shifts its mass towards the aggressive prior. This occurs because the adaptive attacker updates his idealized probability after failed attacks, slowly driving his strategy towards a more aggressive one. Such a shift is visible  through  Figure \ref{fig:attack_freq}, where the frequency of strong attacks increases over time. Consequently, the defender updates her posterior mass towards the aggressive prior (Figure \ref{fig:id_prior_adaptive_main}), yielding more aggressive estimated attack probabilities (Figures \ref{fig:probadapS0} and \ref{fig:probAdapS1}) and deploying substantial defences (Figure \ref{fig:defense_heatmap_adap}).

Finally, observe how these differences in attacker types yield different realized outcomes, with the adaptive attacker achieving greater success in the first stage. Nevertheless, the learning mechanism allows the defender to effectively adapt, switching to a substantial defence in the second stage when necessary, neutralizing both the stationary and the non-stationary attacker in stage 2 (Figures \ref{fig:outcome_freq_fixed} and \ref{fig:outcome_freq_adaptive}).

Overall, these results illustrate how the ARA-FPD framework effectively tracks and anticipates attacker behaviour from the defender's perspective, ultimately optimizing strategic decisions against both stationary and non-stationary opponents.

\section{Conclusions} \label{sec:Conclusion}
This paper has introduced a novel approach to ARA using FPD to model the attacker’s decision problem, employing ideal probabilities rather than explicit utilities. This greatly simplifies ARA assessments and computations. We applied this framework to single-stage interactions, where we obtain an explicit solution, and extended it to multi-stage interactions. For this, we proposed a recursive backward-forward algorithm, alternating between choosing optimal decisions (using the explicit solution) and inducing ideal probabilities in previous stages to be proportional to how close the next stage density would be to his ideals.
Beliefs about the attacker’s ideals are updated using Bayesian learning, allowing the defender to adapt its behaviour based on observed trajectories.

Experiments in drone combat scenarios showcased the approach. The defender was supported in selecting effective defences against both stationary and non stationary attackers. Compared with standard ARA, our ARA-FPD formulation alleviates the computational burden associated with ARA's simulation-optimization approach, albeit the curse of dimensionality remains when the outcome space is large, as we need to induce the ideal probability of each outcome separately. This challenge may be mitigated through 
gradient-based belief updates under dynamic settings, as we shall explore in future work. 

The ARA-FPD approach also shows potential in: 1) proposing new methods to reinforcement learning (RL) problems under threats in providing adversarially robust policies; 2) handling multi-agent reinforcement learning (MARL) scenarios, where each agent could model other agents' beliefs and preferences, mitigating common knowledge assumptions; and, 3) developing much needed explainable methods in MARL, thanks to the explainable nature of goal conditioned learning. These extensions constitute promising avenues that we shall pursue in the future. 



\section*{Acknowledgments}
This work was supported in part by the Spanish Ministry of Science and Innovation program PID2025-172412OB-C21 and the Severo Ochoa Excellence Program CEX-2023-001347-S. MFC is staff hired under the Generation D initiative, promoted by Red.es, an organisation attached to the Ministry for Digital Transformation and the Civil Service, for the attraction and retention of talent through grants and training contracts, financed by the Recovery, Transformation and Resilience Plan through the European Union's Next Generation funds. MK and TG were supported by IAMI project HORIZON-CL3-2023-FCT-01-04: 101168272, and  EU COST Action CA24136.

\newpage
\appendix
\section*{Appendix A}
This appendix contains the detailed proofs of the propositions stated in the main text.

\noindent \textbf{Proof of Proposition 1.}

Let us rewrite problem (\ref{eq:VI_Dkl}) as
\begin{small}
\begin{equation} \label{eq:DKLProof}
    \begin{aligned} 
    \argmin_{\bar{p}\in \Gamma} &\left\{ \int_{\mathcal{A}} \bar {p} (a | d) \left[ \log \left( KL [p_A(\theta \vert a,d) \| p_A^i(\theta \vert a, d)] \right) + \log \left( \frac{\bar {p}(a|d)}{p^i_A(a\vert d)} \right )\right]\dd a \right\} = \\
    \argmin_{\hat{p}\in \Gamma} & \Bigg\{ \int_{\mathcal{A}}\bar {p} (a|d) \log \left (\frac{\bar {p} (a|d)}{p^i_A(a\vert d)  \exp\{-KL [p_A(\theta \vert a,d) \| p_A^i(\theta \vert a, d)]\}} \right ) \dd a  \Bigg\}.
    \end{aligned}
\end{equation}
\end{small}
The solution of (\ref{eq:DKLProof}) remains the same if we add and subtract a constant, say $\log(\mu)$,
\begin{small}
\begin{equation*}
    \begin{aligned}
         & \argmin_{\bar{p}\in \Gamma } \Bigg\{\int_{\mathcal{A}}\bar{p}(a|d) \log \left (\frac{\bar{p}(a|d)}{p^i_A(a\vert d) \exp\{-KL[p_A(\theta \vert a,d) \| p_A^i(\theta \vert a, d)]\}\mu^{-1}} \right )\dd a -\log(\mu)
      \Bigg\}= \\
        & \argmin_{\bar{p}\in \Gamma } \Big\{ KL \Big[ \bar {p}(a|d) \Big\|  p^i_A(a\vert d) \exp\{-KL [p_A(\theta \vert a,d) \| p_A^i(\theta \vert a, d)]\}\mu^{-1}  \Big]\Big\}.
    \end{aligned}
\end{equation*}
\end{small}
Finally, choosing $\mu = Z$, and recalling that, for any two distributions, the KL divergence is
non-negative and  null only when they coincide,
  we find that the explicit solution is given by  (\ref{KL_DivExplicitsol}).
\hfill $\square$

\vspace{0.2cm}
\noindent \textbf{Proof of Proposition 2.}

By definition, we have
\begin{equation*}
    \psi_A(d,a) = \int_{\Theta} u_A(a, \theta, d) p_A(\theta \vert a, d) \dd \theta \qquad \text{and} \qquad \psi_A(a) = \int_{\mathcal{D}} \psi_A(d,a) p_A(d) \dd d.
\end{equation*}

In the sequential case (a), we expand the expected utility $\psi_A(d,a)$ using (\ref{eq:ImplicitUtilityOutcome})
\begin{equation*}
    \begin{aligned}
        \psi_A(d,a) &= \int_{\Theta} \left[ \log p^i_A(\theta \vert a, d) - \log p_A(\theta \vert a, d) \right] p_A(\theta \vert a, d) \dd \theta = - \int_{\Theta} p_A(\theta \vert a, d) \log \left( \frac{p_A(\theta \vert a, d)}{p^i_A(\theta \vert a, d)} \right) \dd \theta \\
        &= -KL[p_A(\theta \vert a, d) \| p^i_A(\theta \vert a, d)].
    \end{aligned}
\end{equation*}
Note that in the FPD formulation the attacker seeks to minimize the outcome divergence $$\argmin_{a \in \mathcal{A}} KL[p_A(\theta \vert a, d) \| p_A^i(\theta \vert a, d)].$$
Since minimising a function $f(x)$ is the same as maximising $-f(x)$, this is equivalent to $$\argmax_{a \in \mathcal{A}} \psi_A(d,a).$$

In the simultaneous case (b), substituting the derived relationship $\psi_A(d,a) = -KL[p_A(\theta \vert a, d) \| p_A^i(\theta \vert a, d)]$ into the simultaneous FPD objective yields
\begin{equation*}
    \argmin_{a \in \mathcal{A}} \int_{\mathcal{D}} KL[p_A(\theta \vert a, d) \| p^i_A(\theta \vert a, d)] p_A(d) \dd d \iff \argmax_{a \in \mathcal{A}} \int_{\mathcal{D}} \psi_A(d,a) p_A(d) \dd d,
\end{equation*}
therefore proving the result.
\hfill $\square$

\vspace{0.2cm}
\noindent \textbf{Proof of Proposition 3.}

Let us first prove the sequential case (a). Expanding the joint KL divergence conditioned on the defender's action $d$:
\begin{equation*}
    KL[p_A(a, \theta \vert d) \| p^i_A(a, \theta \vert d)] = \int_{\mathcal{A}} \int_{\Theta} p_A(a, \theta \vert d) \log \left( \frac{p_A(a, \theta \vert d)}{p^i_A(a, \theta \vert d)} \right) \dd\theta \dd a.
\end{equation*}
Applying the chain rule of probability, $p_A(a, \theta \vert d) = p_A(\theta \vert a, d) p_A(a \vert d)$, the logarithm decomposes into two parts:
\begin{equation*}
    \log \left( \frac{p_A(\theta \vert a, d) p_A(a \vert d)}{p^i_A(\theta \vert a, d) p^i_A(a \vert d)} \right) = \log \left( \frac{p_A(\theta \vert a, d)}{p^i_A(\theta \vert a, d)} \right) + \log p_A(a \vert d) - \log p^i_A(a \vert d).
\end{equation*}
The first term inside the logarithm is exactly the negative of the utility function $u_A(a, \theta, d)$ used in Proposition \ref{thm:HardEquivalence}. Substituting this decomposition back into the integral, we separate the objective
\begin{equation*}
    \begin{aligned}
       KL[p_A(a, \theta \vert d) \| p^i_A(a, \theta \vert d)] &= \int_{\mathcal{A}} p_A(a \vert d) \underbrace{ \left[ \int_{\Theta} -u_A(a, \theta, d) p_A(\theta \vert a, d) \dd\theta \right] }_{= -\psi_A(d,a)} \dd a \\
        &\quad + \int_{\mathcal{A}} p_A(a \vert d) \log p_A(a \vert d) \dd a - \int_{\mathcal{A}} p_A(a \vert d) \log p^i_A(a \vert d) \dd a.
    \end{aligned}
\end{equation*}
Noting that the second term is the negative Shannon entropy, $-\mathcal{H}(p_A(a \vert d))$, the minimization problem becomes
\begin{equation*}
    \argmin_{p_A \in \Gamma} \left[ - \int_{\mathcal{A}} p_A(a \vert d) \Big( \psi_A(d,a) + \log p^i_A(a \vert d) \Big) \dd a - \mathcal{H}(p_A(a \vert d)) \right].
\end{equation*}
Finally, since minimising a function $f(x)$ is equivalent to maximising $-f(x)$, we factor out the negative sign and obtain (\ref{eq:SoftSequential}).

For the simultaneous case (b), expand the formulation
\begin{equation*}
    KL[p_A(a, \theta, d) \| p^i_A(a, \theta, d)] = \int_{\mathcal{A}} \int_{\mathcal{D}} \int_{\Theta} p_A(a, \theta, d) \log \left( \frac{p_A(a, \theta, d)}{p^i_A(a, \theta, d)} \right) \dd\theta \dd d \dd a.
\end{equation*}
Because the interaction is simultaneous, as in previous sections, we assume that the attacker's action is independent of the specific defence, meaning $p_A(a \vert d) = p_A(a)$. Applying the chain rule and the assumption that the defender's strategy is \textit{left to its fate} ($p^i_A(d) = p_A(d)$), the logarithm decomposes into
\begin{equation*}
    \log \left( \frac{p_A(\theta \vert a, d) p_A(a) p_A(d)}{p^i_A(\theta \vert a, d) p^i_A(a) p_A(d)} \right) = \log \left( \frac{p_A(\theta \vert a, d)}{p^i_A(\theta \vert a, d)} \right) + \log p_A(a) - \log p^i_A(a).
\end{equation*}

Substituting this back yields
\begin{equation*}
    \begin{aligned}
         KL[p_A(a, \theta, d) \| p^i_A(a, \theta, d)] &= \int_{\mathcal{A}} p_A(a) \left[ \int_{\mathcal{D}} \underbrace{ \left( \int_{\Theta} -u_A(a, \theta, d) p_A(\theta \vert a, d) \dd\theta \right) }_{= -\psi_A(d,a)} p_A(d) \dd d \right] \dd a \\
        &\quad + \int_{\mathcal{A}} p_A(a) \log p_A(a) \underbrace{ \int_{\mathcal{D}} \int_{\Theta} p_A(\theta \vert a, d) p_A(d) \dd\theta \dd d }_{= 1} \dd a \\
        &\quad - \int_{\mathcal{A}} p_A(a) \log p^i_A(a) \underbrace{ \int_{\mathcal{D}} \int_{\Theta} p_A(\theta \vert a, d) p_A(d) \dd\theta \dd d }_{= 1} \dd a 
        \\ &= \int_{\mathcal{A}} p_A(a) \Big( -\psi_A(a) - \log p^i_A(a) \Big) \dd a - \mathcal{H}(p_A(a)).
    \end{aligned}
\end{equation*}
 Therefore,
\begin{equation*}
    \argmin_{p_A \in \Gamma} KL[p_A(a, \theta, d) \| p^i_A(a, \theta, d)]= \argmin_{p_A \in \Gamma} \left[ \int_{\mathcal{A}} p_A(a) \Big( -\psi_A(a) - \log p^i_A(a) \Big) \dd a - \mathcal{H}(p_A(a)) \right].
\end{equation*}
Converting the minimisation into a maximisation problem, we recover the entropy-regularized objective (\ref{eq:SoftSimultaneous}).
\hfill $\square$
\section*{Appendix B}
This appendix shows how a 2-stage case of the model in Section \ref{sec:Rep_games} would be handled through ARA. Figure \ref{AAA} presents the defender's perspective. Assuming that the utility function $u_D(d_1,d_2,\theta_1,\theta_2)$, and the distributions $p_D(\theta_2|d_2,\theta_1, a_2)$, $p_D(a_2|a_1)$, $p_D(a_1)$, and $p_D(\theta_1|d_1,a_1)$ are available, we proceed as follows to solve $D$'s problem, using \cite{Shachter1986}'s ID algorithm.

     \begin{footnotesize}     
        \begin{enumerate}
            \item Remove chance node $\theta_2$, computing the expected utility
            $\psi_D(d_1,d_2,\theta_1, a_2) = \int u_D(d_1,d_2,\theta_2)p_D(\theta_2|d_2,\theta_1,a_2)\dd\theta_2.$
            \item Remove chance node $A_2$, computing 
             $ 
            \psi_D(d_1,d_2,\theta_1,a_1) = \int \psi_D(d_1,d_2,\theta_1,a_2)p_D(a_2|a_1)\dd a_2. 
            $ 
            \item Remove chance node $\theta_1$, computing  
            $ 
            \psi_D(d_1,d_2,a_1) = \int \psi_D(d_1,d_2,\theta_1,a_1)p_D(\theta_1|a_1,d_1) \dd \theta_1.
            $ 
            \item Remove chance node $A_1$, computing 
            $ 
            \psi_D(d_1,d_2) = \int \psi_D(d_1,d_2,a_1)p_D(a_1)\dd a_1.   
            $ 
            \item Remove decision node $D_2$, computing  
            $ 
            \psi_D(d_1) = \max_{d_2}\psi_D(d_1,d_2).
            $ 
            \item Remove decision node $D_1$, computing 
            $ 
             d_1^{*} = \argmax_{d_1} \psi_D(d_1).   
            $ 
    \end{enumerate}
    \end{footnotesize}
Clearly, we could combine steps 5 and 6, jointly removing nodes $D_1$ and $D_2$, to solve for $$(d_1^{*},d_2^{*}) = \argmax_{(d_1,d_2) \in \mathcal{D}_1 \times \mathcal{D}_2} \psi_D(d_1,d_2).$$ 

\noindent As in Section \ref{sec:Seq_Games}, the required densities $p_D(a_2|a_1)$ and $p_D(a_1)$ include strategic elements. Figure \ref{BBB} portrays $A$'s problem from $D$'s point of view. 
    If we model our uncertainty about its ingredients with a random utility $U_A$ and random densities $P_A$ and $\Pi_A$ for each stage,
\begin{footnotesize}    
$F=(U_A(a_1,a_2,\theta_2),P_A(\theta_2|a_2,\theta_1,d_2),P_A(\theta_1|a_1,d_1),\Pi_A(d_2 \vert d_1), \Pi_A(d_1)) $\end{footnotesize}
    we would solve $A$'s problem through the steps 
    \begin{footnotesize}
    \begin{enumerate}  
        \item Remove chance node $\theta_2$, computing the expected random utility
        $Psi_A(a_1,a_2,\theta_1, d_2) = \int U_A(a_1,a_2,\theta_2)P_A(\theta_2|a_2,\theta_1,d_2)\dd\theta_2. $
            \item Remove chance node $D_2$, computing 
        {  $\Psi_A(a_1,a_2,\theta_1,d_1) =  \int \Psi_A(a_1,a_2,\theta_1,d_2)\Pi_A(d_2|d_1)\dd d_2. $
        }
        \item Remove chance node $\theta_1$, computing  
        $  \Psi_D(a_1,a_2,d_1) =  \int \Psi_D(a_1,a_2,\theta_1,d_1)P_A(\theta_1|a_1,d_1) \dd \theta_1.
        $ 
        \item Remove chance node $D_1$, computing 
        $ \Psi_A(a_1,a_2) = \int \Psi_A(a_1,a_2,d_1)\Pi_A(d_1)\dd d_1. $
        \item Remove decision node $A_2$, computing  
        $  \Psi_A(a_1) = \max_{a_2}\Psi_A(a_1,a_2) $ 
        and  $  A_2^{*} (a_1) = \argmax_{d_2} \Psi_A(a_1, a_2).   $ 
        \item Remove decision node $A_1$, computing 
        $  A_1^{*} = \argmax_{a_1} \Psi_A(a_1).   $ 
\end{enumerate}
\end{footnotesize}
Finally, $D$ obtains $p_D(a_2 | a_1)$, $p(a_1)$ as the densities associated with the distributions $ \mathbb{P}_F[ A^*_2 (a_1, a) \leq a_2 ]$ and $\mathbb{P}_F[ A^*_1 \leq a_1]$ respectively (in general using MC), which serves as input to $D$'s problem. As in the 1-stage case, this approach involves eliciting the attacker's random utility function, which may be hard to obtain. Furthermore, eliciting $\Pi_A(d_1)$ and $\Pi_A(d_2 \vert d_1)$ leads to a recursion of nested decision models, which makes the approach computationally expensive.

\printcredits

\bibliographystyle{cas-model2-names}

\bibliography{references}



\end{document}